\documentclass[12pt]{article}  
\usepackage{graphicx}
\usepackage[a4paper,textwidth=490.0pt,textheight=703.1pt]{geometry}
\usepackage{color}
\usepackage{cite}
\usepackage[rflt]{floatflt}

\usepackage{array}

\usepackage[english]{babel}
\usepackage[latin1]{inputenc}
\usepackage[T1]{fontenc}
\usepackage{ae}

\usepackage{url}

\usepackage{amsmath, amsthm, amssymb}
\newtheorem{thm}{Theorem}[section]

\newtheorem{definition}[thm]{Definition}

\newcommand{\MS}{\overline{{\sf MS}}}

 \newcommand{\HA}{H }

\newcommand{\N}{\nonumber}

\newcommand{\Li}{{\rm Li}}

\newcommand{\ep}{\varepsilon}

\usepackage{rotating}

\usepackage{algorithm}
\usepackage{algorithmicx}
\usepackage{algpseudocode}
\usepackage{graphicx}

\newcounter{mmacnt}
\def\restartmma{\setcounter{mmacnt}{0}}
\restartmma \catcode`|=\active
\def|#1|{\mathrm{#1}}
\catcode`|=12
\newenvironment{mma}{
 \par\smallskip
 \catcode`|=\active
 \parskip=0pt\parindent=0pt 
 \small
 \def\In##1\\{%
   \def\linebreak{\hfill\break\null\qquad}%
   \refstepcounter{mmacnt}
   \hangindent=2.5em\hangafter=0
   \leavevmode
   \llap{\tiny\sffamily In[\arabic{mmacnt}]:=\kern.5em}%
   \mathversion{bold}\footnotesize$\displaystyle##1$\normalsize
   \mathversion{normal}\par
 }%
 \def\Print##1\\{%
   \def\linebreak{\hfill\break}%
   \hangindent=2.5em\hangafter=0
   \leavevmode ##1\par}%
 \def\Out##1\\{%
   \def\linebreak{$\hfill\break\null\hfill$}%
   \kern\abovedisplayskip\par
   \hangindent=2.5em\hangafter=0
   \leavevmode
   \llap{\tiny\sffamily Out[\arabic{mmacnt}]=\kern.5em}
   \footnotesize$\displaystyle##1$\normalsize\hfill\null\par
   \kern\belowdisplayskip
 }%
 \def\Warning##1##2\\{%
   \def\linebreak{\hfill\break}%
   \hangindent=2.5em\hangafter=0
   \leavevmode
   {\scriptsize##1 : ##2}\par}%
}{%
 \par\smallskip
}

\usepackage{color}

\newenvironment{fshaded}{%
\MakeFramed {\FrameRestore}
}%
{\endMakeFramed}

\usepackage{tikz}
\usetikzlibrary{matrix}

\allowdisplaybreaks[4]

\usepackage{hyperref}

\begin{document}
\setlength{\baselineskip}{0.515cm}
\sloppy
\thispagestyle{empty}
\begin{flushleft}
DESY 21-003
\hfill 
\\
DO-TH 21/01 \\
TTP 21-001 \\
RISC Report Series 21-01\\
MSUHEP-21-002\\
SAGEX-21-01 \\
January 2021\\
\end{flushleft}

\mbox{}
\vspace*{\fill}
\begin{center}

{\Large\bf The Polarized Transition Matrix Element \boldmath{$A_{gq}(N)$} of the} 

\vspace*{1mm}
{\Large\bf Variable Flavor Number Scheme at \boldmath{$O(\alpha_s^3)$}}

\vspace{4cm}
\large
A.~Behring$^a$, 
J.~Bl\"umlein$^b$, 
A.~De Freitas$^b$, 
A.~von~Manteuffel$^c$,
K.~Sch\"onwald$^{a,b}$,
and C.~Schneider$^d$ 

\vspace{1.5cm}
\normalsize
{\it $^a$~Institut f\"ur Theoretische Teilchenphysik Campus S\"ud,\\
Karlsruher Institut f\"ur Technologie (KIT) D-76128 Karlsruhe, Germany}
\\

\vspace*{3mm}
{\it  $^b$ Deutsches Elektronen-Synchrotron, DESY,}\\
{\it  Platanenallee 6, D-15738 Zeuthen, Germany}\\

\vspace*{3mm}
{\it $^c$
~Department of Physics and Astronomy,\\
Michigan State University, East Lansing, MI 48824, USA}\\

\vspace*{3mm}
{\it $^d$~Research Institute for Symbolic Computation (RISC),\\
Johannes Kepler University, Altenbergerstra\ss{}e 69,
A-4040, Linz, Austria}


\end{center}
\normalsize
\vspace{\fill}
\begin{abstract}
\noindent 
We calculate the polarized massive operator matrix element $A_{gq}^{(3)}(N)$ to 3-loop order
in Quantum Chromodynamics analytically at general values of the Mellin variable $N$ both in 
the single- and double-mass case in the Larin scheme. It is a transition function required in 
the variable flavor number scheme at $O(\alpha_s^3)$. We also present the results in momentum 
fraction space. 
\end{abstract}

\vspace*{\fill}
\noindent
\numberwithin{equation}{section}
\newpage

\section{Introduction}
\label{sec:1}

\vspace*{1mm}
\noindent
The variable flavor number scheme (VFNS) can be used to translate
twist-2 parton distributions from a scheme with $N_F$ light flavors to a
scheme with $N_F+1$ light flavors at a scale $\mu^2$. Thus, it allows
for a process-independent description of the transition from a massive
quark to a massless quark. In the single heavy mass case this has been
worked out to 2-loop order in \cite{Buza:1996wv} and to 3-loop order in
\cite{Bierenbaum:2009mv}. In terms of the $N_F$-flavor distributions the
new $(N_F+1)$-flavor massless parton densities are given by
\begin{eqnarray}
  {f_k(N_f+1, \mu^2) + f_{\overline{k}}(N_f+1, \mu^2)} 
&=& {A_{qq,Q}^{\rm NS} \Big(N_f, \frac{\mu^2}{m^2}\Big)}\otimes
  {\left[f_k(N_f, \mu^2) + f_{\overline{k}}(N_f, \mu^2)\right]} \nonumber\\
& & + {\tilde A_{qq,Q}^{\rm PS}\Big(N_f, \frac{\mu^2}{m^2}\Big)}\otimes
  {\Sigma(N_f, \mu^2)} 
\nonumber\\ &&
 + {\tilde A_{qg,Q}^{\rm S}\Big(N_f, \frac{\mu^2}{m^2}\Big)}\otimes
  {G(N_f, \mu^2)},
\nonumber\\
  {f_{Q+\bar Q}(N_f+1, \mu^2)} 
   &=& \tilde A_{Qq}^{\rm PS}\Big(N_f, \frac{\mu^2}{m^2}\Big)\otimes
       {\Sigma(N_f, \mu^2)}
   + \tilde A_{Qg}^{\rm S}\Big(N_f, \frac{\mu^2}{m^2}\Big) \otimes
    {G(N_f, \mu^2)},
    \N  \\
 {G(N_f+1, \mu^2)} &=&  {A_{gq,Q}^{\rm S}\Bigl(N_f,\frac{\mu^2}{m^2}\Bigr)}
\otimes {\Sigma(N_f,\mu^2)}
+ A_{gg,Q}^S\Bigl(N_f,\frac{\mu^2}{m^2}\Bigr)\otimes {G(N_f,\mu^2)},  
\N\\
{\Sigma(N_f+1,\mu^2)} &=&
{\sum_{k=1}^{N_f+1} \left[f_k(N_f+1,\mu^2) + 
f_{\overline{k}}(N_f+1,\mu^2)\right]} \nonumber\\
&=& \left[{A_{qq,Q}^{\rm NS}\Big(N_f, \frac{\mu^2}{m^2}\Big)} + 
N_f 
{\tilde{A}_{qq,Q}^{\rm PS} \Big(N_f, \frac{\mu^2}{m^2}\Big)} +
\tilde{A}_{Qq}^{\rm PS} \Big(N_f, \frac{\mu^2}{m^2}\Big) \right]
\nonumber\\ &&
\otimes {\Sigma(N_f,\mu^2)} \nonumber\\ &&
+\left[N_f 
{\tilde{A}^{\rm S}_{qg,Q}\Big(N_f, \frac{\mu^2}{m^2}\Big)} 
+ \tilde{A}^{\rm S}_{Qg}\Big(N_f, \frac{\mu^2}{m^2}\Big) \right]
\otimes {G(N_f,\mu^2)} .
\label{eq:VFNS}
\end{eqnarray}
The quark and antiquark parton densities are denoted by $f_k$ and
$f_{\bar{k}}$, respectively, and $G(N_F,\mu^2)$ is the gluon density.
We write $\Sigma(N_F,\mu^2) = \sum_{k=1}^{N_F} (f_k + f_{\bar{k}})$ for
the singlet-quark density. The massive operator matrix elements (OMEs)
$A_{ij}(N_F,m^2/\mu^2)$ are process-independent quantities and have an
expansion in the strong coupling constant $a_s = \alpha_s/(4\pi)$,
\begin{eqnarray}
A_{ij}(N) = \delta_{ij} + \sum_{k=1}^\infty a_s^k A_{ij}^{(k)}(N).
\end{eqnarray}
Here, $\mu$ denotes the decoupling scale and $m$ is the mass of the
decoupling heavy-quark flavor $Q$. The OMEs explicitly depend on the
mass through logarithms. In total, there are seven different OMEs
contributing to the matching relations. Moreover, we introduce the
shorthand notations
\begin{align}
  \tilde{f}(N_F) &= \frac{f}{N_F}, &
  \hat{f}(N_F) &= f(N_F+1) - f(N_F).
\end{align}
The VFNS is important for obtaining parton distribution functions at
very large virtualities as they are required for, e.g., scattering
processes at the Large Hadron Collider (LHC), the Tevatron, RHIC and the EIC, 
in particular for precision measurements of observables in Quantum
Chromodynamics (QCD) and the determination of the strong coupling
constant $\alpha_s(M_Z)$ \cite{Bethke:2011tr}.
The relations (\ref{eq:VFNS}) apply structurally to both the unpolarized
and polarized case.

The corresponding relations have to be generalized in the case of two
heavy-quark contributions, as for charm and bottom quarks, the masses of
which are very similar,
cf.~\cite{Ablinger:2017err,Blumlein:2018jfm,PVFNS}.
We also provide the 2-mass 3-loop OME $A_{gq}^{(3),\sf two-mass}$, for
which a closed form expression can be obtained \cite{Schonwald:2019gmn}.

The individual OMEs start contributing at different orders: The OME
$\tilde A_{Qg}^{\rm S}$ starts already at 1-loop order, while
$\tilde A_{Qq}^{\rm PS}$, $A_{qq,Q}^{\rm NS}$, $A_{gg,Q}^S$ and
$A_{gq,Q}^{\rm S}$ only contribute from 2-loop order onward. At 3-loop
order, also $\tilde A_{qq,Q}^{\rm PS}$ and $\tilde A_{qg,Q}^{\rm S}$
appear. The complete 2-loop corrections were calculated in
Refs.~\cite{Buza:1995ie,Buza:1996wv,Bierenbaum:2007qe,
Bierenbaum:2009zt,Buza:1996xr,Bierenbaum:2007pn,POLl,PVFNS} for the
unpolarized and polarized cases. Mellin moments for the OMEs at 3-loop
order were calculated in \cite{Bierenbaum:2009mv,MELLM} and for
transversity $A_{qq,Q}^{\rm NS,TR}$ in \cite{Blumlein:2009rg}. Depending
on the process, the moments reached up to $N=13$ at most.

In \cite{Ablinger:2010ty}, the unpolarized OMEs
$\tilde A_{qq,Q}^{\rm PS}(N)$ and $\tilde A_{qg,Q}^{\rm S}(N)$, as well
as the $O(N_f T_F^2 C_{A,F})$ corrections to the OMEs
$\tilde A_{Qg}^{\rm S}$, $\tilde A_{Qq}^{\rm PS}$, $A_{qq,Q}^{\rm NS}$
and $A_{qq,Q}^{\rm NS,TR}$ have been calculated at 3-loop order. Here,
$T_F = 1/2$, $C_A = N_c$, $C_F = (N_c^2-1)/(2 N_c)$ denote the color
factors for the gauge group $SU(N_c)$. For $A_{gg,Q}^S$ and
$A_{gq,Q}^{\rm S}$ the corresponding contributions were calculated in
\cite{Blumlein:2012vq}.
Furthermore, the unpolarized 3-loop OMEs $A_{qq,Q}^{\rm NS}$, 
$\tilde A_{Qq}^{\rm PS}$, $A_{gq,Q}^{\rm S}$ and  $A_{gg,Q}^{\rm S}$
were obtained in
\cite{Ablinger:2014vwa,Ablinger:2014nga,Ablinger:2014lka,Ablinger:2014uka,GLUE}
and the logarithmic contributions to 3-loop order in
Ref.~\cite{Behring:2014eya}. Partial results for $\tilde A_{Qg}^{(3)}$
were calculated in \cite{Ablinger:2017ptf}. The massive 3-loop
polarized OMEs are known in the non-singlet and pure-singlet cases
\cite{Ablinger:2014vwa,Ablinger:2019etw}. The two-mass contributions up
to 3-loop order are known in the unpolarized case
\cite{Ablinger:2017err,Blumlein:2018jfm,Ablinger:2017xml,Ablinger:2018brx}
and in the polarized case in
\cite{Ablinger:2019gpu,Ablinger:2020snj}, in both cases
up to $\tilde A_{Qg}^{(3)}$. Heavy-flavor contributions to charged
current processes up to 3-loop order and the polarized structure
function $g_1^{\rm NS}(x,Q^2)$ were dealt with in
\cite{Blumlein:2014fqa,Behring:2016hpa,Behring:2015roa,Behring:2015zaa}.

In the present paper we compute the complete polarized OME
$A_{gq,Q}^{\rm S, (3)}(N)$ for general values of $N$ in both the single-
and two-mass cases. Note that we will drop the superscript ${\rm S}$ in
the following, since for this OME only the singlet part contributes.
Due to the crossing relations, cf. \cite{Blumlein:1996vs}, only the odd
moments contribute and they are used to construct the analytic
continuation to complex values of $N$ or the Bjorken $x$-space,
respectively. From the $O(1/\ep)$ pole term one obtains the contribution
to the 3-loop polarized anomalous dimension $\gamma_{gq}^{(2)} \propto
T_F$, cf.~\cite{Moch:2014sna,Behring:2019tus}. We perform the
calculation using the Larin scheme \cite{Larin:1993tq}, which is a
consistent scheme w.r.t. the $\gamma_5$ problem, see also
\cite{Behring:2019tus}. It is convenient to work in this scheme also for
the description of the observables in polarized deep-inelastic
scattering. This requires that also the polarized parton distribution
functions are evolved using this scheme and that one calculates the
massless Wilson coefficients in this scheme. Observables, such as the
deep-inelastic structure functions, are then scheme-independent. The
OME $A_{gq,Q}^{(3)}(N)$ requires to use a special projector for the
external quark lines, which has first been derived in
Ref.~\cite{Behring:2019tus}, Eq.~(11).

The paper is organized as follows. We discuss technical details of the
calculation in Section~\ref{sec:2}. In Section~\ref{sec:3}, we present
the constant part of the unrenormalized single-mass 3-loop OME
$A_{gq,Q}^{(3)}$ in Mellin-$N$ space and discuss its small- and
large-$x$ behavior. Section~\ref{sec:4} is devoted to the analytic
calculation of the OME $A_{gq,Q}^{(3), \sf two-mass}$.
Section~\ref{sec:5} contains the conclusions. In the Appendix, we
present the polarized single- and two-mass OME $A_{gq,Q}^{(3)}$ in
Mellin- and momentum-fraction space, treating the heavy-quark mass in
both in the on-shell and $\MS$ scheme.

\section{The Formalism}
\label{sec:2}

\vspace*{1mm}
\noindent
Concerning the formalism, we follow closely
Ref.~\cite{Ablinger:2014lka}, in which the corresponding result in the
unpolarized case has been calculated.
Renormalizing the heavy-quark mass in the on-shell scheme and the
coupling constant in the $\MS$ scheme, the massive operator matrix
element $A_{gq,Q}^{(3)}$ has the structure \cite{Bierenbaum:2009mv}
\begin{eqnarray}
    A_{gq,Q}^{(3)}&=&
                     -\frac{\gamma_{gq}^{(0)}}{24}
                      \Biggl\{
                          \gamma_{gq}^{(0)}\hat{\gamma}_{qg}^{(0)}
                         +\Bigl(
                              \gamma_{qq}^{(0)}
                             -\gamma_{gg}^{(0)}
                             +10\beta_0
                             +24\beta_{0,Q}
                                     \Bigr)\beta_{0,Q}
                      \Biggr\}
                           \ln^3 \Bigl(\frac{m^2}{\mu^2}\Bigr)
                     +\frac{1}{8}\Biggl\{
                         6 {\gamma_{gq}^{(1)}}\beta_{0,Q}
\N\\ &&
                        +\hat{\gamma}_{gq}^{(1)}\Bigl(
                                      \gamma_{gg}^{(0)}
                                     -\gamma_{qq}^{(0)}
                                     -4\beta_0
                                     -6\beta_{0,Q}
                                                 \Bigr)
                        +\gamma_{gq}^{(0)}\Bigl(
                                       \hat{\gamma}_{qq}^{(1), {\sf NS}}
                                      +\hat{\gamma}_{qq}^{(1), {\sf PS}}
                                      -\hat{\gamma}_{gg}^{(1)}
                                      +2\beta_{1,Q}
                                                 \Bigr)
                      \Biggr\}
                           \ln^2 \Bigl(\frac{m^2}{\mu^2}\Bigr)
\N\\ &&
                     +\frac{1}{8}\Biggl\{
                              4 {\hat{\gamma}_{gq}^{(2)}}
                            + 4a_{gq,Q}^{(2)}         \Bigl(
                                    \gamma_{gg}^{(0)}
                                   -\gamma_{qq}^{(0)}
                                   -4\beta_0
                                   -6\beta_{0,Q}
                                                       \Bigr)
                            + 4\gamma_{gq}^{(0)}       \Bigl(
                                      a_{qq,Q}^{(2),{\sf NS}}
                                     +a_{Qq}^{(2),{\sf PS}}
                                     -a_{gg,Q}^{(2)}
\N\\ &&
                                     +\beta_{1,Q}^{(1)}
                                                       \Bigr)
                            + \gamma_{gq}^{(0)}\zeta_2 \Bigl(
                               \gamma_{gq}^{(0)}\hat{\gamma}_{qg}^{(0)}
                               +\Bigl[
                                        \gamma_{qq}^{(0)}
                                       -\gamma_{gg}^{(0)}
                                       +12\beta_{0,Q}
                                       +10\beta_0
                                            \Bigr]\beta_{0,Q}
                                                       \Bigr)
                      \Biggr\}
                           \ln \Bigl(\frac{m^2}{\mu^2}\Bigr)
\N\\ &&
                  + \overline{a}_{gq,Q}^{(2)} \Bigl(
                                       \gamma_{qq}^{(0)}
                                      -\gamma_{gg}^{(0)}
                                      +4\beta_0
                                      +6\beta_{0,Q}
                                             \Bigr)
                  + \gamma_{gq}^{(0)} \Bigl(
                                       \overline{a}_{gg,Q}^{(2)}
                                      -\overline{a}_{Qq}^{(2),{\sf PS}}
                                      -\overline{a}_{qq,Q}^{(2),{\sf NS}}
                                             \Bigr)
                -\gamma_{gq}^{(0)}\beta_{1,Q}^{(2)}
\N\\ && 
                -\frac{\gamma_{gq}^{(0)}\zeta_3}{24} \Bigl(
                           \gamma_{gq}^{(0)}\hat{\gamma}_{qg}^{(0)}
                          +\Bigl[
                                   \gamma_{qq}^{(0)}
                                  -\gamma_{gg}^{(0)}
                                  +10\beta_0
                                      \Bigr]\beta_{0,Q}
                                             \Bigr)
                -\frac{3\gamma_{gq}^{(1)}\beta_{0,Q}\zeta_2}{8}
                +2 \delta m_1^{(-1)} a_{gq,Q}^{(2)}
\N\\ &&
                +\delta m_1^{(0)} \hat{\gamma}_{gq}^{(1)}
                +4 \delta m_1^{(1)} \beta_{0,Q} \gamma_{gq}^{(0)}
                +a_{gq,Q}^{(3)}~.
\label{AgqQ3Ren1}
\end{eqnarray}
This expression depends on the Riemann $\zeta$-function evaluated at
integer values, $\zeta_k = \sum_{l=1}^\infty l^{-k},~k \in \mathbb{N},~k \geq 2$,
the polarized anomalous dimensions $\gamma_{ij}^{(k)}$
up to three-loop order (i.e. $k=0,1,2$) \cite{Moch:2014sna,Behring:2019tus},
the expansion coefficients of the QCD $\beta$-function
\cite{Tarasov:1980au,Larin:1993tp,vanRitbergen:1997va,Czakon:2004bu,
Baikov:2016tgj,Herzog:2017ohr,Luthe:2017ttg},
terms from mass renormalization \cite{MASS}
and the constant parts of the unrenormalized massive OMEs $a_{ij}^{(k)}$
in the polarized case \cite{Buza:1996xr,Bierenbaum:2007pn,POLl,PVFNS},
again up to three-loop order.
The expansion coefficients related to the QCD $\beta$-function are given by
\begin{eqnarray}
        \beta_0 &=& \frac{11}{3}C_A - \frac{4}{3}T_F N_F~,
        \\
        \beta_{0,Q} &=& -\frac{4}{3}T_F~,
        \\
        \beta_{1,Q} &=& -4 \left( \frac{5}{3}C_A +C_F \right) T_F~,
        \\
        \beta_{1,Q}^{(1)} &=& -\frac{32}{9}T_F C_A + 15 T_F C_F~,
        \\
        \beta_{1,Q}^{(2)} &=& -\frac{86}{27}T_FC_A-\frac{31}{4}T_FC_F
                              -T_F\left(\frac{5}{3}C_A+C_F\right)\zeta_2  ~,
\end{eqnarray}
and the terms stemming from mass renormalization read
\begin{eqnarray}
        \delta m_{1}^{(-1)} &=& 6 C_F~,
        \\
        \delta m_{1}^{(0)} &=& -4 C_F~,
        \\
        \delta m_{1}^{(1)} &=& C_F \left( 4 + \frac{3}{4}\zeta_2 \right)~.
\end{eqnarray}
From the logarithmic terms one can extract the complete 2-loop anomalous dimension
$\gamma_{gq}^{(1)}$ \cite{Mertig:1995ny,Vogelsang:1995vh,Moch:2014sna,Behring:2019tus}
and the contributions $\propto T_F$ of $\gamma_{gq}^{(2)}(N)$, one
of the anomalous dimensions at 3-loop order \cite{Moch:2014sna,Behring:2019tus}.

We use a well established approach to calculate the 86 contributing Feynman diagrams.
First the diagrams are generated with an extension of {\tt QGRAF} \cite{Nogueira:1991ex}
which can deal with local operator insertions \cite{Bierenbaum:2009mv}.
The Feynman rules are then inserted in {\tt TFORM} \cite{Tentyukov:2007mu}
where also the Dirac- and color-traces are calculated.
The local operator insertions are resummed into generating functions
using the auxiliary variable $t$, cf.~\cite{Ablinger:2012qm}.
This introduces on top of the usual denominators from particle propagators
denominators which depend linearly on the loop momenta and the variable $t$.
The scalar integrals are subsequently reduced to a minimal set of master integrals
using the implementation of integration-by-parts reduction \cite{IBP}
in {\tt Reduze2} \cite{vonManteuffel:2012np}, which can also deal
with linear propagators.
The solutions of the master integrals are obtained using standard 
techniques.\footnote{For a recent review, see Ref.~\cite{Blumlein:2018cms}.}
This includes methods based on hypergeometric functions 
\cite{GHYP,Slater,Appell,Hamberg,Bierenbaum:2007qe,Ablinger:2012qm}, 
Mellin-Barnes representations \cite{MELB} 
and differential equations \cite{Ablinger:2015tua,Ablinger:2018zwz}.
For the analytic continuation of Mellin-Barnes integrals, the 
packages {\tt MB} \cite{Czakon:2005rk} and {\tt MBresolve} \cite{Smirnov:2009up}
were used.
When applying methods based on direct integration and Mellin-Barnes 
representations the results are typically given by multiple
sums over hypergeometric expressions which can still depend
on the dimensional parameter $\varepsilon=D-4$.
These expressions can be expanded in $\varepsilon$ and
the resulting sums can 
afterwards be performed utilizing modern summation technology
\cite{Karr:81,Schneider:01,Schneider:05a,Schneider:07d,Schneider:08c,
Schneider:10a,Schneider:10b,Schneider:10c,Schneider:13b}
as encoded in the packages {\tt Sigma} \cite{SIG1,SIG2}, {\tt 
HarmonicSums} \cite{HARMONICSUMS,Ablinger:2010kw,Ablinger:2013hcp}, 
{\tt EvaluateMultiSums}, {\tt SumProduction} \cite{EMSSP}, 
and {\tt $\rho$-Sum} \cite{RHOSUM}.
For one of the master integrals it was essential to apply
the multivariate Almkvist-Zeilberger algorithm \cite{AZ}
as implemented in the package {\tt MultiIntegrate} \cite{Ablinger:2013hcp,Ablinger:2015tua}
on the Mellin-space representation of the master integral.
This way we were able to directly compute a difference equation 
for the Mellin-space result which we solved using the same 
summation technology cited before.
The final results for individual master integrals,
diagrams and the full final result 
have been checked by computing a number of
integer moments with {\tt MATAD} \cite{Steinhauser:2000ry}.

As in the unpolarized case, the polarized OME $A_{gq,Q}^{(3)}$
can be completely expressed by harmonic sums $S_{\vec{a}}(N)$
and $\zeta$-values \cite{Blumlein:2009cf} in Mellin-space and harmonic polylogarithms
$H_{\vec{a}}(x)$ and $\zeta$-values in Bjorken $x$-space.
The definitions of harmonic sums and harmonic polylogarithms
are given by the iterative formulas \cite{HSUM}
\begin{eqnarray}
        S_{b,\vec{a}}(N) = \sum_{k=1}^N \frac{{\rm sign}(b)^k}{k^{|b|}} S_{\vec{a}}(k),~S_{\emptyset}(N) = 1;~a_i,b \in 
        \mathbb{Z} \setminus \{0\},~N \in \mathbb{N} \setminus \{0\}
\end{eqnarray}
and \cite{Remiddi:1999ew}
\begin{eqnarray}
        H_{b,\vec{a}}(x) = \int\limits_{0}^{1}dx f_{b}(x) H_{\vec{a}}(x),~H_{\emptyset}(x) = 1,~
        H_{\underbrace{0,\dots,0}_{\text{n times}}} (x) = \frac{1}{n!}\ln^n(x)
\end{eqnarray}
with $a \in \{-1,0,1\}$ and 
\begin{eqnarray}
        f_1(x) = \frac{1}{1-x},~ f_0(x) = \frac{1}{x},~ f_{-1}(x) = \frac{1}{1+x} ~.
\end{eqnarray}

The full renormalization and mass factorization of all massive 
operator matrix elements including $A_{gq,Q}^{(3)}$ up to 
3-loop order has been presented in Ref.~\cite{Bierenbaum:2009mv} 
for the single mass case and in Ref.~\cite{Ablinger:2017err} 
for the two-mass case.
The necessary steps are the renormalization of the masses,
the coupling constant and the twist-2 light cone operators.
Furthermore, collinear singularities have to be removed 
by mass factorization. 
Contrary to the massless case, the $Z$-factors related to the 
ultraviolet renormalization in the massive case are
not inverse to those describing the collinear singularities.
Moreover, the coupling constant is first renormalized in
a MOM scheme using the background-field method \cite{BGF}
and afterwards translated to the usual $\MS$ scheme in 
order to fulfill the on-shell condition of the external
partonic states.

\section{The Single-Mass Correction}
\label{sec:3}

\vspace*{1mm}
\noindent
In the single-mass case, the renormalized OME~(\ref{AgqQ3Ren1}) can be
expressed in terms of lower-order terms as well as the newly evaluated
constant part $a_{gq}^{(3)}(N)$ of the unrenormalized OME. We define
\begin{eqnarray}
\bar{p}_{gq}  = \frac{2 + N}{N (1 + N)}
\end{eqnarray}
and use the shorthand notation $S_{\vec{k}}(N) \equiv S_{\vec{k}}$.
One obtains
\begin{eqnarray}
        a_{gq}^{(3)}(N) &=& \frac{1}{2} [1 - (-1)^N] 
        \Biggl\{
                \textcolor{blue}{C_F T_F^2}
                \Biggl\{
                        N_F \bar{p}_{gq}
                        \biggl[
                                \frac{32 \big(616+2109 N+2334 N^2+868 N^3\big)}{243 (1+N)^3}
                                \nonumber \\ &&
                                +\biggl(
                                        -\frac{32 \big(28+55 N+30 N^2\big)}{27 (1+N)^2}
                                        -\frac{16}{9} S_2
                                \biggr) S_1
                                +\frac{16 (2+5 N)}{27 (1+N)} S_1^2
                                -\frac{16}{27} S_1^3
                                \nonumber \\ &&
                                +\frac{16 (2+5 N)}{27 (1+N)} S_2
                                -\frac{32}{27} S_3
                                +\biggl(
                                        \frac{16 (2+5 N)}{9 (1+N)}
                                        -\frac{16}{3} S_1
                                \biggr) \zeta_2
                                -\frac{224}{9} \zeta_3
                        \biggr]
                        \nonumber \\ &&
                        +\bar{p}_{gq}
                        \biggl[
                                \frac{16 \big(157+957 N+1299 N^2+607 N^3\big)}{243 (1+N)^3}
                                -\biggl(
                                         \frac{32 \big(25+48 N+29 N^2\big)}{27 (1+N)^2}
                                        +\frac{32}{9} S_2
                                \biggr) S_1
                                \nonumber \\ &&
                                +\frac{32 (2+5 N)}{27 (1+N)}  S_1^2
                                -\frac{32}{27} S_1^3
                                +\frac{32 (2+5 N)}{27 (1+N)}  S_2
                                -\frac{64}{27} S_3
                                +\biggl(
                                        \frac{32 (2+5 N)}{9 (1+N)}
                                        -\frac{32}{3} S_1
                                \biggr) \zeta_2
                                \nonumber \\ &&
                                +\frac{512}{9} \zeta_3
                        \biggr]
                \Biggr\}
                + \textcolor{blue}{C_F^2 T_F}
                \Biggl\{
                         \frac{4 \zeta_3 P_{11}}{9 (N-1) N^3 (1+N)^3}
                        +\frac{8 P_{12}}{81 (N-1) N^3 (1+N)^3} S_3
                        \nonumber \\ &&
                        -\frac{4 P_{14}}{81 (N-1) N^4 (1+N)^4} S_2
                        -\frac{4 P_{16}}{243 (N-1)^2 N^5 (1+N)^5 (2+N)}
                        +\bar{p}_{gq}
                        \biggl[
                                \frac{64}{3} B_4
                                \nonumber \\ &&
                                +\biggl(
                                        -\frac{8 P_{10}}{243 N^3 (1+N)^3}
                                        +\frac{4 (125+167 N)}{27 (1+N)}  S_2
                                        -\frac{208}{27} S_3
                                        +\frac{32}{9} S_{2,1}
                                \biggr) S_1
                                \nonumber \\ &&
                                +\biggl(
                                        \frac{4 P_5}{81 N^2 (1+N)^2}
                                        -\frac{28}{9} S_2
                                \biggr) S_1^2
                                -\frac{4 \big(48+119 N+197 N^2\big)}{81 N (1+N)}  S_1^3
                                +\frac{58}{27} S_1^4
                                \nonumber \\ &&
                                -\frac{2}{3} S_2^2
                                -\frac{220}{9} S_4
                                -\frac{16 (23+29 N)}{27 (1+N)} S_{2,1}
                                +\frac{32}{3} S_{3,1}
                                -\frac{32}{9} S_{2,1,1}
                                -\big(
                                         \frac{2 P_4}{9 N^2 (1+N)^2}
                                         \nonumber \\ &&
                                        +\frac{4 \big(12+5 N+11 N^2\big)}{9 N (1+N)} S_1
                                        -\frac{20}{3} S_1^2
                                        +\frac{28}{3} S_2
                                \biggr) \zeta_2
                                -\frac{192}{5} \zeta_2^2
                                +\frac{304}{9} S_1 \zeta_3
                        \biggr]
                        \nonumber \\ &&
                        -\frac{128}{(N-1) N^2 (1+N)^2} S_{-1} S_2
                        -\biggl(
                                 \frac{128 (3+N) (1+2 N)}{3 (N-1)^2 N (1+N)^3 (2+N)}
                                 \nonumber \\ &&
                                -\frac{128}{(N-1) N^2 (1+N)^2}  S_{-1}
                        \biggr) S_{-2}
                        -\frac{128}{3 (N-1) N^2 (1+N)^2} S_{-3}
                        \nonumber \\ &&
                        +\frac{128}{(N-1) N^2 (1+N)^2}  S_{2,-1}
                        -\frac{128}{(N-1) N^2 (1+N)^2}  S_{-2,-1}
                \Biggr\}
                \nonumber \\ &&
                + \textcolor{blue}{C_A C_F T_F} 
                \Biggl\{
                         \frac{16 P_1}{27 (N-1) N^2 (1+N)^2}  S_{-3}
                        -\frac{4 P_2}{9 (N-1) N^2 (1+N)^2} \zeta_3 
                        \nonumber \\ &&
                        +\frac{16 P_3}{81 (N-1) N^2 (1+N)^2} S_3
                        -\frac{4 P_7}{27 (N-1) N^2 (1+N)^3} S_2
                        -\frac{4 P_8}{81 N^3 (1+N)^3}  S_1^2
                        \nonumber \\ &&
                        -\frac{4 P_{15}}{243 (N-1)^2 N^5 (1+N)^5 (2+N)}
                        +\bar{p}_{gq}
                        \biggl[
                                -\frac{32}{3} B_4
                                +\biggl(
                                        -\frac{848}{27} S_3
                                        -\frac{64}{9} S_{2,1}
                                        \nonumber \\ &&
                                        +32 S_{-2,1}
                                \biggr) S_1
                                -\frac{304}{9} S_{-3} S_1
                                -\frac{58}{27} S_1^4
                                -\frac{116}{9} S_1^2 S_2
                                -\frac{122}{9} S_2^2
                                -\frac{356}{9} S_4
                                \nonumber \\ &&
                                -\biggl(
                                         \frac{176}{9} S_1^2
                                        +\frac{176}{9} S_2
                                \biggr) S_{-2}
                                -\frac{32}{9} S_{-2}^2
                                -\frac{320}{9} S_{-4}
                                +\frac{32}{3} S_{3,1}
                                +\frac{224}{9} S_{-2,2}
                                +\frac{272}{9} S_{-3,1}
                                \nonumber \\ &&
                                +16 S_{2,1,1}
                                -\frac{352}{9} S_{-2,1,1}
                                +\biggl(
                                        -\frac{20}{3} S_1^2
                                        -4 S_2
                                        -8 S_{-2}
                                \biggr) \zeta_2
                                +\frac{192}{5} \zeta_2^2
                                -\frac{112}{9} S_1 \zeta_3
                        \biggr]
                        \nonumber \\ &&
                        +\biggl(
                                \frac{8 P_{13}}{243 N^4 (1+N)^4}
                                -\frac{4 \big(12+194 N-61 N^2-253 N^3\big)}{27 (N-1) N^2 (1+N)}  S_2
                        \biggr) S_1
                        \nonumber \\ &&
                        +\frac{4 \big(204+262 N+711 N^2+293 N^3\big)}{81 N^2 (1+N)^2} S_1^3
                        +\frac{64}{(N-1) N^2 (1+N)^2} S_{-1} S_2
                        \nonumber \\ &&
                        -\biggl(
                                 \frac{16 P_9}{27 (N-1)^2 N (1+N)^3 (2+N)}
                                +\frac{32 \big(29+28 N-71 N^2-40 N^3\big)}{27 (N-1) N (1+N)^2} S_1
                                \nonumber \\ &&
                                +\frac{64}{(N-1) N^2 (1+N)^2} S_{-1}
                        \biggr) S_{-2}
                        -\frac{16 \big(48-27 N+11 N^2-14 N^3\big)}{9 (N-1) N^2 (1+N)^2} S_{2,1}
                        \nonumber \\ &&
                        -\frac{64}{(N-1) N^2 (1+N)^2} S_{2,-1}
                        +\frac{32 \big(1+2 N-37 N^2-20 N^3\big)}{27 (N-1) N (1+N)^2} S_{-2,1}
                        \nonumber \\ &&
                        +\frac{64}{(N-1) N^2 (1+N)^2} S_{-2,-1}
                        -\biggl(
                                 \frac{4 P_6}{9 N^3 (1+N)^3}
                                 \nonumber \\ &&
                                -\frac{4 \big(60+70 N+135 N^2+53 N^3\big)}{9 N^2 (1+N)^2} S_1
                        \biggr) \zeta_2
                \Biggr\}
        \Biggr\}
        ,
\end{eqnarray}
with the polynomials
\begin{eqnarray}
        P_1 &=& 40 N^4+83 N^3-22 N^2-11 N+36,
        \\
        P_2 &=& 89 N^4+370 N^3-169 N^2+30 N-608,
        \\
        P_3 &=& 136 N^4+152 N^3-43 N^2-53 N-138,
        \\
        P_4 &=& 204 N^4+390 N^3+187 N^2+37 N+114,
        \\
        P_5 &=& 697 N^4+1283 N^3+736 N^2+60 N+72,
        \\
        P_6 &=& 7 N^5-5 N^4-9 N^3+29 N^2-100 N-12,
        \\
        P_7 &=& 231 N^5+408 N^4+77 N^3-602 N^2-1202 N+8,
        \\
        P_8 &=& 1141 N^5+3817 N^4+4142 N^3+2708 N^2-396 N-288,
        \\
        P_9 &=& 106 N^6+389 N^5+96 N^4-920 N^3-800 N^2+27 N+238,
        \\
        P_{10} &=& 230 N^6+1179 N^5+2481 N^4+2354 N^3+1074 N^2+198 N+108,
        \\
        P_{11} &=& 281 N^6+891 N^5+423 N^4-799 N^3-1112 N^2-500 N+240,
        \\
        P_{12} &=& 511 N^6+1431 N^5+457 N^4-1131 N^3-428 N^2+240 N+648,
        \\
        P_{13} &=& 4307 N^7+19468 N^6+33504 N^5+31031 N^4+11038 N^3+1608 N^2-1440 N
        \nonumber \\ &&
        -432,
        \\
        P_{14} &=& 2207 N^8+8327 N^7+8423 N^6-451 N^5-5122 N^4-4636 N^3-4860 N^2
        \nonumber \\ &&
        +1296,
        \\
        P_{15} &=& 7027 N^{12}+39120 N^{11}+73621 N^{10}+17722 N^9-143181 N^8-181350 N^7
        \nonumber \\ &&
        +17183 N^6+97038 N^5+11306 N^4-53746 N^3-5916 N^2-2808 N-432,
        \\
        P_{16} &=& 14748 N^{12}+83610 N^{11}+133975 N^{10}-53587 N^9-315078 N^8-143766 N^7
        \nonumber \\ &&
        +221994 N^6+176898 N^5-29869 N^4-10811 N^3+44106 N^2+684 N
        \nonumber \\ &&
        +1512.
\end{eqnarray}
The color factors in QCD take on the values $C_A = 3, C_F = 4/3, T_F = 1/2$.
The nested sums and constants in $a_{gq}^{(3)}(N)$ have weights up to
$\mathsf{w}=4$ and the constant
\begin{eqnarray}
B_4 &=&-4\zeta_2\ln^2(2) +\frac{2}{3}\ln^4(2)
-\frac{13}{2}\zeta_4 +16 {\rm Li}_4\Bigl(\frac{1}{2}\Bigr)
\label{eqB4} 
\end{eqnarray}
appears. Here, we write $\Li_k(x) = \sum_{l=1}^\infty x^l/l^k,~~|x| \leq
1$ for the classical polylogarithm. As a cross-check, we compared this
result to an independent calculation of the moments $N=3,5,7$ using
\texttt{MATAD} \cite{Steinhauser:2000ry} and we find agreement.

The nested sums can be mapped to a basis using algebraic reduction
\cite{Blumlein:2003gb} after which only the sums
\begin{eqnarray}
S_1, S_2, S_3, S_4, S_{-1}, S_{-2}, S_{-3}, S_{-4}, S_{2,1}, S_{2,-1}, S_{-2,-1}, S_{-2,1},
S_{-2,2}, S_{3,1}, S_{-3,1}, S_{2,1,1}, S_{-2,1,1}
\end{eqnarray}
appear.
In addition, structural relations, such as multiple argument relations
and differentiation, \cite{Blumlein:2009ta}, can be applied, which
leaves us only with    
\begin{eqnarray}
S_1, S_{2,1}, S_{-2,1}, S_{-3,1}, S_{2,1,1}, S_{-2,1,1}
\end{eqnarray}
as basic sums.
At $N=1$, the OME $a_{gq}^{(3)}(N)$ has a removable singularity. In this
limit the expression becomes
\begin{eqnarray}
a_{gq}^{(3)}(N \rightarrow 1) &=&
\textcolor{blue}{C_F T_F} \Biggl\{
        \textcolor{blue}{T_F} \Biggl[
                -\frac{508}{81}
                +\frac{8}{3} \zeta_2
                +\frac{256}{3} \zeta_3
                +\textcolor{blue}{N_F} 
                \Biggl(
                         \frac{7858}{81}
                        +\frac{4 \zeta_2}{3}
                        -\frac{112 \zeta_3}{3}
                \Biggr)
        \Biggr]
\nonumber\\ &&
        +\textcolor{blue}{C_F} \Biggl[
                -\frac{26665}{54}
                +48 B_4
                -91 \zeta_2
                -\frac{432 \zeta_2^2}{5}
                +364 \zeta_3
        \Biggr]
        +\textcolor{blue}{C_A} \Biggl[
                 \frac{11705}{162}
                 \nonumber\\ &&
                -24 B_4
                +\frac{109 \zeta_2}{3}
                +\frac{432 \zeta_2^2}{5}
                -\frac{682 \zeta_3}{3}
        \Biggr]
\Biggr\}. 
\end{eqnarray}
This agrees with the expectation that the rightmost singularity for
gluonic OMEs occurs at $N=0$. As was observed in
\cite{Blumlein:2013ota}, removable singularities can also appear at
rational values of $N>0$ for massive OMEs. The OME $A_{gq}^{(3)}(N)$ is
a meromorphic function \cite{Blumlein:2009ta} since it can be expressed
in terms of harmonic sums \cite{HSUM} over $\mathbb{Q}(N)$ with rational
weights whose denominators factorize, with factors of the form
$(N-k)^l, k \in \mathbb{Z}, l \in \mathbb{N}$.
The poles of this OME are located at negative integers, $N \leq 0$. 

We now turn our discussion to the behavior of $a_{gq}^{(3)}(N)$ in the
limits $N \to \infty$ and $N \to 0$, i.e. the `leading singularity' in
the polarized case.

Around $N \to \infty$ the constant part of the OME the asymptotic
behavior is given by
\begin{eqnarray}
a_{gq}^{(3)}(N \rightarrow \infty)  &=&
\textcolor{blue}{C_F T_F} \Biggl\{
        -\frac{58}{27} (\textcolor{blue}{C_A - C_F}) \frac{L^4(N)}{N}
        + \biggl[
                \frac{1172}{81} \textcolor{blue}{C_A}
                -\frac{788}{81} \textcolor{blue}{C_F} 
                \nonumber\\ &&
                - \frac{16}{27} \bigl( 2 + \textcolor{blue}{N_F} \bigr) \textcolor{blue}{T_F} 
        \biggr] \frac{L^3(N)}{N}
        +\biggl[
                \frac{80}{27} (2+\textcolor{blue}{N_F}) \textcolor{blue}{T_F}
                +\textcolor{blue}{C_F} \left(
                        \frac{2788}{81}
                        +\frac{32 \zeta_2}{9}
                \right)
                \nonumber\\ &&
                -\textcolor{blue}{C_A} \left(
                        \frac{4564}{81}
                        +\frac{88 \zeta_2}{9}
                \right)
        \biggr] \frac{L^2(N)}{N}
        +\biggl[
                -\textcolor{blue}{T_F} 
                \left(
                         \frac{928}{27}
                        +\frac{128 \zeta_2}{9}
                        +\textcolor{blue}{N_F} \left(
                                \frac{320}{9}
                                +\frac{64 \zeta_2}{9}
                        \right)
                \right)
                \nonumber\\ &&
                -\textcolor{blue}{C_F} \left(
                        \frac{1840}{243}
                        -\frac{536 \zeta_2}{27}
                        -\frac{896 \zeta_3}{27}
                \right) 
                +\textcolor{blue}{C_A} \left(
                         \frac{34456}{243}
                        +\frac{112 \zeta_2}{3}
                        -\frac{1424 \zeta_3}{27}
                \right)
        \biggr] \frac{L(N)}{N}
        \nonumber\\ &&
\Biggr\} + O\left(\frac{1}{N}\right),
\end{eqnarray}
where we abbreviate ${L}(N) = \ln(N) + \gamma_E$ and we write $\gamma_E$
for the Euler-Mascheroni constant. The $N \to \infty$ limit in $N$-space
corresponds to the $x \to 1$ limit in $x$-space. This allows us to
deduce that in this limit the leading singular term is $\propto
\alpha_s^3 \ln^4(1-x)$.

The position of the so-called `leading-poles' can be inferred from an
analysis of the anomalous dimensions of different scattering processes
in fixed-order perturbation theory. For massless vector operators they
are located at $N=1$ \cite{Gross:1974cs}, for massless quark operators
at $N=0$ \cite{Kirschner:1983di,Blumlein:1995jp} and for massless scalar
operators at $N=-1$ \cite{Blumlein:1998mg}.
The leading term of $a_{gq}^{(3)}(N)$ in an expansion around $N=0$ reads
\begin{eqnarray}
a_{gq}^{(3)}(N \rightarrow 0)  \hspace*{-2mm} &=& 
\textcolor{blue}{C_F T_F} 
\Biggl\{
         \left[
                 \frac{32}{9} \textcolor{blue}{C_A} 
                -\frac{112}{9} \textcolor{blue}{C_F}
        \right] \frac{1}{N^5}
        +\left[
                 \frac{32}{3} \textcolor{blue}{C_A} 
                +\frac{1024}{27} \textcolor{blue}{C_F}
        \right] \frac{1}{N^4}
        +\left[
                \textcolor{blue}{C_A} 
                \left(
                         \frac{200}{81}
                        -\frac{80 \zeta_2}{9}
                \right) \right.
\nonumber\\ && \left.
                -\textcolor{blue}{C_F} 
                \left(
                        \frac{37636}{81}
                        +\frac{520 \zeta_2}{9}
                        -\frac{64 \zeta_3}{3}
                \right)
        \right] \frac{1}{N^3}
        + \Biggl[
                \textcolor{blue}{C_F} 
                \left(
                        \frac{380152}{243}
                        +\frac{3296 \zeta_2}{27}
                        -\frac{768 \zeta_2^2}{5} 
                \right. 
                \nonumber\\ &&  \left.
                        +\frac{176 \zeta_3}{3}
                \right)
                +\textcolor{blue}{C_A} 
                \left(
                         \frac{102800}{243}
                        +\frac{1024 \zeta_2}{27}
                        -256 \zeta_3
                \right)
        \Biggr] \frac{1}{N^2}
\nonumber\\ &&                
+        \Biggl[
                \textcolor{blue}{T_F} 
                \left(
                         \frac{5024}{243}
                        +\frac{128}{9} \zeta_2
                        +\frac{1024}{9} \zeta_3
                        +\textcolor{blue}{N_F} 
                        \left(
                                 \frac{39424}{243}
                                +\frac{64 \zeta_2}{9}
                                -\frac{448 \zeta_3}{9}
                        \right)
                \right)
\nonumber\\ &&
                -\textcolor{blue}{C_F} 
                \left(
                        \frac{11972}{3}
                        +\frac{64}{3} B_4
                        +\frac{28510}{81} \zeta_2
                        -\frac{18016}{45} \zeta_2^2
                        -\frac{736}{27} \zeta_3
                        -640 \zeta_5
                \right)
\nonumber\\ &&                       
                -\textcolor{blue}{C_A} 
                \left(
                        \frac{365768}{243} 
                        -\frac{32}{3} B_4
                        +\frac{1124 \zeta_2}{81}
                        -\frac{1264 \zeta_2^2}{9}
                        -\frac{2488 \zeta_3}{9}
                \right)
        \Biggr] \frac{1}{N}
\Biggr\} + O\left(N^0\right).
\end{eqnarray}
The leading behavior in $x$-space is $\propto \alpha_s^3 \ln^4(1/x)$,
but the coefficients of the sub--leading terms have an oscillating sign
while their magnitude increases with increasing logarithmic order, which
strongly compensates the leading term in the physical region which is
relevant, for example, at the EIC \cite{Boer:2011fh}. This is in line
with earlier observations in other cases, cf.
Refs.~\cite{Blumlein:1995jp,Blumlein:1996hb,Blumlein:1997em,Blumlein:1998mg}.
For the complete OME $A_{gq,Q}^{(3)}$ in $N$ and $x$-space we refer to
the Appendix.

\section{The Two-Mass Correction}
\label{sec:4}

\vspace*{1mm}
\noindent
The 2-loop two-mass OME has been calculated in \cite{PVFNS}. The 3-loop two-mass OME 
$A_{gq}^{(3),\sf two-mass}$ is calculated as follows. Using the projector for the external quark lines
given in \cite{Behring:2019tus} the following representation is obtained
\begin{align}
         \hat{A}_{gq,Q}^{(3),\sf two-mass} &= \textcolor{blue}{C_F T_F^2} \left[ 384\frac{d-6}{d-2} I_{gq,Q}(N) 
        + \frac{1536}{d-2} I_{gq,Q}(N-1) + \bigl( m_1 \leftrightarrow m_2 \bigr) \right],
\end{align}
where the function $I_{gq,Q}(N)$ is given by
\begin{align}
        I_{gq,Q}(N) &= \left( \frac{4\pi}{\mu} \right)^{3\ep/2} \frac{ \Gamma(6-3d/2) \Gamma(N+1)}{\Gamma(N+d/2)} 
        \int\limits_0^1 dz_1 \int\limits_0^1 dz_2 \int\limits_0^1 dz_3
        \left[z_1(1-z_1)\right]^{1+\ep/2} \left[z_2(1-z_2)\right]^{1+\ep/2} 
\nonumber \\ &
        \times \left[z_3(1-z_3)\right]^{-1-\ep/2} \left[ \frac{z_3 m_1^2}{z_1 (1-z_1)} + \frac{(1-z_3)m_2^2}{z_2(1-z_2)} \right]^{3\ep/2}.
\end{align}
The $N$-dependence completely factorizes from the dependence of the masses.
The integral can be performed with analytic Mellin-Barnes integral techniques. 
We define the mass ratio by
\begin{eqnarray}
\eta = \frac{m_2^2}{m_1^2} < 1 
\end{eqnarray}
and
\begin{eqnarray}
L_1 = \ln\left(\frac{m_1^2}{\mu^2}\right),~~~~~~
L_2 = \ln\left(\frac{m_2^2}{\mu^2}\right).
\end{eqnarray}
For the unrenormalized OME 
one obtains
\begin{align}
        \hat{A}_{gq,Q}^{(3),\sf two-mass} &= 
        \textcolor{blue}{C_F T_F^2} (N+2) S_\ep^3 
        \Biggl\{
         \frac{1024}{9 N (N+1) \ep^3}
        +\frac{1}{\ep^2}
        \biggl[
                \frac{256}{3 N (N+1)} ( L_1 + L_2)
                +\frac{512 (2+5 N)}{27 N (N+1)^2}
\nonumber \\ &
                -\frac{512 }{9 N (N+1)} S_1
        \biggr]
        +\frac{1}{\ep}
        \biggl[
                \frac{64 }{N (N+1)} ( L_1^2 + L_2^2 )
                + \biggl(
                        \frac{128 (2+5 N)}{9 N (1+N)^2}
                        +\frac{64}{N (1+N)} \HA_0(\eta )
\nonumber \\ &
                        -\frac{128}{3 N (1+N)} S_1
                \biggr) L_1
                + \biggl(
                        \frac{128 (2+5 N)}{9 N (1+N)^2}
                        -\frac{64 }{N (1+N)} \HA_0(\eta )
                        -\frac{128 }{3 N (1+N)} S_1
                \biggr) L_2
\nonumber \\ &
                +\frac{128 }{3 N (1+N)} \HA_0^2(\eta )
                +\frac{128 \big(25+48 N+29 N^2\big)}{27 N (1+N)^3}
                -\frac{256 (2+5 N)}{27 N (1+N)^2} S_1
                +\frac{128 }{9 N (1+N)} S_1^2
\nonumber \\ &
                +\frac{128 }{9 N (1+N)} S_2
                +\frac{128 }{3 N (1+N)} \zeta_2
        \biggr]
        \Biggr\} + a_{gq}^{(3),\sf two-mass}
\\
a_{gq}^{(3),\sf two-mass} &= \textcolor{blue}{C_F T_F^2} (N+2) 
        \Biggl\{
                 \frac{32 }{N (1+N)} ( L_1^3 + L_2^3 )
                + \biggl(
                        \frac{32 (2+5 N)}{3 N (1+N)^2}
\nonumber \\ &
                        +\frac{48 }{N (1+N)} \HA_0(\eta )
                        -\frac{32 }{N (1+N)} S_1
                \biggr) L_1^2
                +\biggl(
                        \frac{32 (2+5 N)}{3 N (1+N)^2}
                        -\frac{48 }{N (1+N)} \HA_0(\eta )
\nonumber \\ &
                        -\frac{32 }{N (1+N)} S_1
                \biggr) L_2^2
                +\biggl(
                        \frac{32 \big(25+48 N+29 N^2\big)}{9 N (1+N)^3}
                        +\frac{32 (2+5 N)}{3 N (1+N)^2} \HA_0(\eta )
                        \nonumber \\ &
                        +\frac{32 }{N (1+N)} \HA_0^2(\eta )
                        +\biggl\{
                                -\frac{64 (2+5 N)}{9 N (1+N)^2}
                                -\frac{32 }{N (1+N)} \HA_0(\eta )
                        \biggr\} S_1
                        +\frac{32 }{3 N (1+N)} S_1^2
                        \nonumber \\ &
                        +\frac{32 }{3 N (1+N)} S_2
                        +\frac{32 }{N (1+N)} \zeta_2
                \biggr) L_1
                +\biggl(
                        \frac{32 \big(25+48 N+29 N^2\big)}{9 N (1+N)^3}
                        \nonumber \\ &
                        -\frac{32 (2+5 N)}{3 N (1+N)^2} \HA_0(\eta )
                        +\frac{32 }{N (1+N)} \HA_0^2(\eta )
                        +\biggl\{
                                -\frac{64 (2+5 N)}{9 N (1+N)^2}
                                \nonumber \\ &
                                +\frac{32 }{N (1+N)} \HA_0(\eta )
                        \biggr\} S_1
                        +\frac{32 }{3 N (1+N)} S_1^2
                        +\frac{32 }{3 N (1+N)} S_2
                        +\frac{32 }{N (1+N)} \zeta_2
                \biggr) L_2
                \nonumber \\ &
                +\frac{20 (1-\eta^2) }{3 \eta  N (1+N)} \HA_0(\eta )
                -\frac{T_{1}}{3 \eta  N (1+N)^2} \HA_0^2(\eta )
                -\frac{16 }{9 N (1+N)} \HA_0^3(\eta )
\nonumber \\ &
                -\frac{32 }{3 N (1+N)} \HA_0^2(\eta ) \HA_1(\eta )
                +\frac{64 }{3 N (1+N)} \HA_0(\eta ) \HA_{0,1}(\eta )
                -\frac{64 }{3 N (1+N)} \HA_{0,0,1}(\eta )
\nonumber \\ &
                -\frac{8 T_{2}}{243 \eta  N (1+N)^4}
                +\frac{(1+\eta ) \big(5+22 \eta +5 \eta ^2\big)}{6 \eta ^{3/2} N (1+N)}
                \biggl[
                         \HA_0^2(\eta) \left\{ \HA_{-1}\big(\sqrt{\eta }\big) + \HA_1\big(\sqrt{\eta }\big) \right\}
\nonumber \\ &
                        - 4 \HA_0(\eta) \left\{ \HA_{0,-1}\big(\sqrt{\eta }\big) + \HA_{0,1}\big(\sqrt{\eta }\big) \right\}
                        + 8 \left\{ \HA_{0,0,-1}\big(\sqrt{\eta }\big) + \HA_{0,0,1}\big(\sqrt{\eta }\big) \right\}
                \biggr]
\nonumber \\ &
                - \biggl[
                         \frac{64 \big(25+48 N+29 N^2\big)}{27 N (1+N)^3}
                        +\frac{64 }{3 N (1+N)} \HA_0^2(\eta )
                        +\frac{64 }{9 N (1+N)} S_2
                \biggr] S_1
                \nonumber \\ &
                +\frac{64 (2+5 N)}{27 N (1+N)^2} S_1^2
                -\frac{64 }{27 N (1+N)} S_1^3
                +\frac{64 (2+5 N)}{27 N (1+N)^2} S_2
                -\frac{128 }{27 N (1+N)} S_3
                \nonumber \\ &
                +\biggl[
                        \frac{64 (2+5 N)}{9 N (1+N)^2}
                        -\frac{64}{3 N (1+N)} S_1
                \biggr] \zeta_2
                -\frac{128 }{9 N (1+N)} \zeta_3
        \Biggr\}
        ,
\end{align}
with $S_\ep = \exp[\tfrac{\ep}{2}(\gamma_E - \ln(4\pi))]$ and the polynomials
\begin{align}
T_{1} &= 5 \eta ^2 N+5 \eta ^2-78 \eta  N-14 \eta +5 N+5,
\\
T_{2} &= 405 \eta ^2 N^3+1215 \eta ^2 N^2+1215 \eta ^2 N+405 \eta ^2-3238 \eta  N^3-7626 \eta  N^2-6258 \eta  
N-1438 \eta 
\nonumber \\ &
	+405 N^3+1215 N^2+1215 N+405 .
\end{align}
Since the color-factor contribution of $O(C_F T_F^2)$ does not receive a finite renormalization, it is 
directly given in the $M$ scheme \cite{Matiounine:1998re,Moch:2014sna}.
We have checked the results using {\tt q2e/exp} \cite{q2e} by calculating the moments $N=3,5$  
expanding in the first powers of $\eta$, cf. also~\cite{Ablinger:2017err}.
Note that by expanding the OME in powers of $\eta$ the root-structures in the above expressions disappear 
showing a dependence on $\eta$ only.

We can recover the $\mathcal{O}(T_F^2)$ part of the single mass OME $\hat{A}_{gq,Q}$
by performing the limit $\eta \to 1$.
For the $\mathcal{O}(\ep^0)$ part one obtains
\begin{align}
        a_{gq,Q}^{(3),T_F^2} &=
        \textcolor{blue}{C_F T_F^2} \frac{\bar{p}_{gq}}{2} 
\Biggl\{
        \frac{32 \big(157+957 N+1299 N^2+607 N^3\big)}{243 (N+1)^3}
        -64 \biggl[
                 \frac{25+48 N+29 N^2}{27 (N+1)^2}
\nonumber \\ &
                +\frac{64}{9} S_2
        \biggr] S_1
        +\frac{64 (2+5 N)}{27  (N+1)} S_1^2
        -\frac{64}{27} S_1^3
        +\frac{64 (2+5 N)}{27 (N+1)} S_2
\nonumber \\ &
        -\frac{128}{27} S_3
        +64 \biggl[
                \frac{2+5 N}{9 (N+1)}
                -\frac{1}{3} S_1
        \biggr] \zeta_2
        +\frac{1024}{9} \zeta_3
\Biggr\},
\end{align}
which agrees with the result given before.
The renormalized two-mass OME in $N$ and $x$-space is given in the Appendix.

\section{Conclusions}
\label{sec:5}

\vspace*{1mm}
\noindent
In this paper we have calculated the single and two-mass contributions
to the massive operator matrix element $A_{gq}^{(3)}(N)$,
which contributes to the matching relations of the VFNS at 3-loop order.
On the technical side of the calculation, we have used the the
integration-by-parts program {\tt Reduze2} to reduce the scalar
integrals with local operator insertions to a minimal set of master
integrals. The master integrals have been computed using different
techniques based on generating functions.
These techniques allowed us to find difference equations for the Mellin
space results, which were subsequently solved with the packages 
{\tt Sigma}, {\tt EvaluateMultiSums}, {\tt SumProduction},
{\tt $\rho$-Sum} and {\tt HarmonicSums}.
As in the unpolarized case, the polarized matrix element
$A_{gq}^{(3)}(N)$ can be expressed in terms of harmonic sums up to
weight {\sf w = 4} in Mellin space and harmonic polylogarithms up to
weight {\sf w = 5} in Bjorken $x$-space.
For the two-mass relation the dependence on the Mellin variable $N$ and
the squared mass ratio $\eta$ factorize.
Note that other massive operator matrix elements
\cite{Ablinger:2014nga,Ablinger:2019gpu,Ablinger:2014uka,GLUE}
also depend on more complicated structures like generalized harmonic
sums and finite binomial sums.
As in the unpolarized case \cite{Ablinger:2014lka}, diagrams of the
Benz topology contribute. Additionally, we presented the results for the
renormalization of the heavy-quark mass in the on-shell and
$\overline{\sf MS}$ scheme.
 
\appendix
\section{Appendix}
\label{sec:A}

\vspace*{1mm}
\noindent
The massive OME $A_{gq,Q}(N)$ with the strong coupling constant renormalized in the 
$\MS$ scheme and the heavy-quark mass in the on-shell scheme obeys the following
expansion up to 3-loop order:
\begin{eqnarray}
A_{gq,Q}(N, a_s) &=&    a_s^2 A_{gq,Q}^{(2)}(N)
                        + a_s^3 A_{gq,Q}^{(3)}(N)
\end{eqnarray}
with $a_s = \alpha_s^{\MS}(\mu^2)/(4\pi)$. 
The 2-loop OME $A_{gq,Q}^{(2)}(N)$ is 
given in Ref.~\cite{PVFNS}. The 3-loop OME reads
\begin{eqnarray}
A_{gq,Q}^{(3)}(N) &=& \frac{1}{2}\left[1-(-1)^N\right]
\textcolor{blue}{C_F T_F} 
\Biggl\{
        L_M^3 
        \biggl[
                \textcolor{blue}{C_F} \bar{p}_{gq} 
                \biggl(
                        \frac{4 \big(6+N+N^2\big)\big(-4+3 N+3 N^2\big)}{9 N^2 (1+N)^2}
                        \nonumber \\ &&
                        -\frac{16}{9} S_1
                \biggr)
                +\textcolor{blue}{C_A} \bar{p}_{gq} 
                \biggl(
                        -\frac{8 \big(4+11 N+11 N^2\big)}{9 N (1+N)}
                        +\frac{16}{9} S_1
                \biggr)
                +\frac{32}{9} \bar{p}_{gq} \textcolor{blue}{T_F} (2+\textcolor{blue}{N_F}) 
        \biggr]
        \nonumber \\ &&
        +L_M^2 
        \biggl[
                \textcolor{blue}{C_A} 
                \biggl(
                        \frac{4 Q_{10}}{9 N^3 (1+N)^3}
                        +\bar{p}_{gq} 
                        \biggl(
                                -\frac{8}{3} S_1^2
                                -8 S_2
                                -16 S_{-2}
                        \biggr)
                        \nonumber \\ &&
                        +\frac{16 \big(6-8 N+9 N^2+5 N^3\big)}{9 N^2 (1+N)^2} S_1
                \biggr)
                +\textcolor{blue}{C_F} \bar{p}_{gq} 
                \biggl(
                        -\frac{2 Q_4}{9 N^2 (1+N)^2}
                        -\frac{16 (3-4 N)}{9 N} S_1
                        \nonumber \\ &&
                        +\frac{8}{3} S_1^2
                        -\frac{40}{3} S_2
                \biggr)
                +\bar{p}_{gq} \textcolor{blue}{T_F} 
                \biggl(
                        \frac{32 (2+5 N)}{9 (1+N)}
                        -\frac{32}{3} S_1
                \biggr)
        \biggr]
        +L_M 
        \biggl[
                \bar{p}_{gq} \textcolor{blue}{T_F} 
                \biggl(
                        \frac{992}{27}
                        \nonumber \\ &&
                        +\textcolor{blue}{N_F} 
                        \biggl(
                                \frac{32 \big(40+83 N+34 N^2\big)}{27 (1+N)^2}
                                +\frac{32 (2+5 N)}{9 (1+N)} S_1
                                -\frac{16}{3} S_1^2
                                -\frac{16}{3} S_2
                        \biggr)
                \biggr)
                \nonumber \\ &&
                +\textcolor{blue}{C_A} 
                \biggl(
                        -\frac{32 Q_1}{3 (N-1) N^2 (1+N)^2} S_{-2}
                        -\frac{8 Q_{11}}{27 N^3 (1+N)^3} S_1
                        \nonumber \\ &&
                        -\frac{8 Q_{13}}{27 (N-1) N^4 (1+N)^4}
                        +\bar{p}_{gq} 
                        \biggl(
                                \frac{128}{3} S_{-2} S_1
                                +\frac{8}{9} S_1^3
                                +\frac{56}{3} S_1 S_2
                                +\frac{256}{9} S_3
                                +\frac{64}{3} S_{-3}
                                \nonumber \\ &&
                                -\frac{64}{3} S_{-2,1}
                                +64 \zeta_3
                        \biggr)
                        -\frac{4 \big(12-110 N-63 N^2-13 N^3\big)}{9 N^2 (1+N)^2} S_1^2
                        \nonumber \\ &&
                        -\frac{4 \big(20-2 N+15 N^2+9 N^3\big)}{3 N^2 (1+N)^2} S_2
                \biggr)
                +\textcolor{blue}{C_F} 
                \biggl(
                        -\frac{2 Q_{15}}{27 (N-1) N^5 (1+N)^4}
                        \nonumber \\ &&
                        +\bar{p}_{gq} 
                        \biggl(
                                -\frac{4 Q_5}{9 N^2 (1+N)^2} S_2
                                +\biggl(
                                        \frac{8 Q_3}{27 N^2 (1+N)^2}
                                        +\frac{88}{3} S_2
                                \biggr) S_1
                                \nonumber \\ &&
                                +\frac{4 \big(12-7 N-25 N^2\big)}{9 N (1+N)} S_1^2
                                -\frac{8}{9} S_1^3
                                +\frac{176}{9} S_3
                                -\frac{32}{3} S_{2,1}
                                -64 \zeta_3
                        \biggr)
                        \nonumber \\ &&
                        +\frac{128}{(N-1) N^2 (1+N)^2} S_{-2}
                \biggr)
        \biggr]
        +\textcolor{blue}{C_A} 
        \biggl(
                \bar{p}_{gq} 
                \biggl[
                        \frac{4 Q_{14}}{81 N^4 (1+N)^4}
                        +\biggl(
                                -\frac{8 Q_8}{81 N (1+N)^3}
                                \nonumber \\ &&
                                -\frac{4 \big(24+41 N+53 N^2\big)}{9 N (1+N)} S_2
                                +\frac{32}{9} S_3
                        \biggr) S_1
                        +\biggl(
                                \frac{4 \big(48+371 N+568 N^2+389 N^3\big)}{27 N (1+N)^2}
                                \nonumber \\ &&
                                +\frac{16}{3} S_2
                        \biggr) S_1^2
                        -\frac{4 \big(8+19 N+31 N^2\big)}{9 N (1+N)} S_1^3
                        +\frac{16}{9} S_1^4
                        \nonumber \\ &&
                        +\frac{4 \big(16+59 N+68 N^2+49 N^3\big)}{9 N (1+N)^2} S_2
                        -\frac{8 \big(8+11 N+11 N^2\big)}{9 N (1+N)} S_3
                        \nonumber \\ &&
                        +\biggl(
                                \frac{20}{3} S_1^2
                                +4 S_2
                                +8 S_{-2}
                        \biggr) \zeta_2
                        +\biggl(
                                \frac{8 \big(4+11 N+11 N^2\big)}{9 N (1+N)}
                                -\frac{16}{9} S_1
                        \biggr) \zeta_3
                \biggr]
                \nonumber \\ &&
                +\biggl(
                        \frac{4 Q_9}{9 N^3 (1+N)^3}
                        -\frac{4 \big(60+70 N+135 N^2+53 N^3\big)}{9 N^2 (1+N)^2} S_1
                \biggr) \zeta_2
        \biggr)
        \nonumber \\ &&
        +\bar{p}_{gq} \textcolor{blue}{T_F} 
        \biggl[
                -\frac{32 \big(98+369 N+408 N^2+164 N^3\big)}{81 (1+N)^3}
                \nonumber \\ &&
                +\textcolor{blue}{N_F} 
                \biggl(
                        -\frac{32 \big(98+369 N+408 N^2+164 N^3\big)}{81 (1+N)^3}
                        +\biggl(
                                \frac{32 \big(22+41 N+28 N^2\big)}{27 (1+N)^2}
                                \nonumber \\ &&
                                +\frac{16}{3} S_2
                        \biggr) S_1
                        -\frac{16 (2+5 N)}{9 (1+N)} S_1^2
                        +\frac{16}{9} S_1^3
                        -\frac{16 (2+5 N)}{9 (1+N)} S_2
                        +\frac{32}{9} S_3
                        \nonumber \\ &&
                        +\biggl(
                                -\frac{16 (2+5 N)}{9 (1+N)}
                                +\frac{16}{3} S_1
                        \biggr) \zeta_2
                        -\frac{32}{9} \zeta_3
                \biggr)
                +\biggl(
                        \frac{32 \big(22+41 N+28 N^2\big)}{27 (1+N)^2}
                        \nonumber \\ &&
                        +\frac{16}{3} S_2
                \biggr) S_1
                -\frac{16 (2+5 N)}{9 (1+N)} S_1^2
                +\frac{16}{9} S_1^3
                -\frac{16 (2+5 N)}{9 (1+N)} S_2
                +\frac{32}{9} S_3
                \nonumber \\ &&
                +\biggl(
                        -\frac{32 (2+5 N)}{9 (1+N)}
                        +\frac{32}{3} S_1
                \biggr) \zeta_2
                -\frac{64}{9} \zeta_3
        \biggr]
        +\textcolor{blue}{C_F} \bar{p}_{gq} 
        \biggl[
                -\frac{8 Q_2}{9 N^2 (1+N)^2} S_3
                \nonumber \\ &&
                +\frac{4 Q_{12}}{27 N^3 (1+N)^3} S_2
                +\frac{Q_{16}}{162 N^4 (1+N)^5}
                +\biggl(
                        \frac{8 Q_7}{81 N (1+N)^3}
                        -\frac{32}{9} S_3
                        \nonumber \\ &&
                        +\frac{4 \big(6+17 N+29 N^2\big)}{9 N (1+N)} S_2
                \biggr) S_1
                +\biggl(
                        -\frac{4 \big(12+116 N+175 N^2+161 N^3\big)}{27 N (1+N)^2}
                        \nonumber \\ &&
                        -\frac{16}{3} S_2
                \biggr) S_1^2
                +\frac{4 \big(2+11 N+23 N^2\big)}{9 N (1+N)} S_1^3
                -\frac{16}{9} S_1^4
                +\frac{16}{3} S_4
                +\biggl(
                        \frac{2 Q_6}{9 N^2 (1+N)^2}
                        \nonumber \\ &&
                        +\frac{4 \big(12+5 N+11 N^2\big)}{9 N (1+N)} S_1
                        -\frac{20}{3} S_1^2
                        +\frac{28}{3} S_2
                \biggr) \zeta_2
                +\biggl(
                        \frac{16}{9} S_1
                        \nonumber \\ &&
                        -\frac{4 \big(6+N+N^2\big)\big(-4+3 N+3 N^2\big)}{9 N^2 (1+N)^2}
                \biggr) \zeta_3
        \biggr]
\Biggr\}
+ a_{gq}^{(3)},
\end{eqnarray}
where we define
\begin{eqnarray}
L_M = \ln\left(\frac{m^2}{\mu^2}\right).
\end{eqnarray}
The polynomials $Q_i$ read
\begin{eqnarray}
Q_1 &=& 5 N^4+9 N^3-4 N^2-4 N+6,
\\
Q_2 &=& 7 N^4+14 N^3+23 N^2+16 N-36,
\\
Q_3 &=& 23 N^4-149 N^3-88 N^2-6 N-36,
\\
Q_4 &=& 69 N^4+66 N^3+43 N^2+46 N+96,
\\
Q_5 &=& 145 N^4+248 N^3+79 N^2-24 N+72,
\\
Q_6 &=& 204 N^4+390 N^3+187 N^2+37 N+114,
\\
Q_7 &=& 472 N^4+1269 N^3+1551 N^2+724 N+132,
\\
Q_8 &=& 1252 N^4+2997 N^3+3360 N^2+1819 N+528,
\\
Q_9 &=& 7 N^5-5 N^4-9 N^3+29 N^2-100 N-12,
\\
Q_{10} &=& 69 N^5+276 N^4+263 N^3+12 N^2+172 N+48,
\\
Q_{11} &=& 197 N^5+791 N^4+952 N^3+148 N^2+348 N+144,
\\
Q_{12} &=& 175 N^6+552 N^5+657 N^4+376 N^3+204 N^2+540 N
+216,
\\
Q_{13} &=& 8 N^8+341 N^7+1276 N^6+617 N^5-1835 N^4+44 N^3+1037 N^2+204 N
\nonumber\\ &&
+36,
\\
Q_{14} &=& 3347 N^8+11540 N^7+16090 N^6+10202 N^5+3200 N^4
+430 N^3+3 N^2
\nonumber\\ &&
+36 N+108,
\\
Q_{15} &=& 51 N^9-300 N^8-674 N^7-360 N^6-1775 N^5+456 N^4-662 N^3-4296 N^2
\nonumber\\ &&
-216 N
+864,
\\
Q_{16} &=& 2067 N^9+12639 N^8+23134 N^7+12958 N^6+2319 N^5+691 N^4+448 N^3
\nonumber\\ &&
+23136 N^2
+15840 N+4752.
\end{eqnarray}
The methods to obtain the analytic continuation of harmonic sums to complex values of $N$ 
are presented in Refs.~\cite{Blumlein:2000hw,Blumlein:2005jg,Blumlein:2009ta,Blumlein:2009fz}.

For the Bjorken $x$-space representation it is convenient to define 
\begin{eqnarray}
p_{gq}(x) = \frac{1}{x}\left[1-(1-x)^2\right]~.
\end{eqnarray}
The massive operator matrix element $A_{gq,Q}^{(3)}(x)$ in Bjorken $x$-space
is given in terms of harmonic polylogarithms \cite{Remiddi:1999ew}.
To shorten the expressions we use 
$H_{\vec{a}}(x) \equiv H_{\vec{a}}$
as a shorthand notation. 
The full expression is given by:
\begin{eqnarray}
A_{gq,Q}^{(3)}(x) &=& 
\textcolor{blue}{C_F T_F}
\Biggl\{
L_M^3 
\biggl[
        \frac{32}{9} p_{gq} \textcolor{blue}{T_F} (2+\textcolor{blue}{N_F}) 
        +\textcolor{blue}{C_F} 
        \biggl(
                 p_{gq} 
                \biggl(
                        -\frac{16}{3} H_0^2
                        -\frac{16}{9} H_1
                \biggr)
                +\frac{4}{3} (-84+85 x)
                \nonumber \\ &&
                -\frac{8}{9} (74+53 x) H_0
        \biggr)
        +\textcolor{blue}{C_A} 
        \biggl(
                -\frac{8}{9} (10+x)
                +\frac{16}{9} (4+x) H_0
                +\frac{16}{9} p_{gq} H_1
        \biggr)
\biggr]
\nonumber \\ &&
+L_M^2 
\Biggl[
        \frac{32}{9} \textcolor{blue}{T_F} (4+x)
        -\frac{32}{3} p_{gq} \textcolor{blue}{T_F} H_1
        +\textcolor{blue}{C_F} 
        \biggl(
                 p_{gq} 
                \biggl(
                        16 H_0 H_1
                        +\frac{8}{3} H_1^2
                        -\frac{32}{3} H_{0,1}
                        \nonumber \\ &&
                        -\frac{16}{3} \zeta_2
                \biggr)
                -\frac{2}{3} (240-217 x)
                -\frac{4}{9} (194+161 x) H_0
                -\frac{8}{3} (8-x) H_0^2
                \nonumber \\ &&
                +\frac{16}{9} (11-7 x) H_1
        \biggr)
        +\textcolor{blue}{C_A} 
        \biggl(
                 p_{gq} 
                \biggl(
                        \frac{16}{3} H_0 H_1
                        -\frac{8}{3} H_1^2
                \biggr)
                -\frac{4}{3} (72-95 x)
                \nonumber \\ &&
                -\biggl(
                         \frac{8}{9} (14+113 x)
                        -16 (2+x) H_{-1}
                \biggr) H_0
                +\frac{32}{3} (1+x) H_0^2
                -\frac{80}{9} (4-5 x) H_1
                \nonumber \\ &&
                -\frac{64}{3} (1+x) H_{0,1}
                -16 (2+x) H_{0,-1}
                +\frac{16}{3} (8+5 x) \zeta_2
        \biggr)
\Biggr]
\nonumber \\ &&
+L_M 
\biggl[
        \textcolor{blue}{C_F} 
        \biggl(
                 p_{gq} 
                \biggl(
                        \frac{4}{3} H_0^4
                        -\frac{16}{3} H_0^2 H_1
                        -\frac{8}{9} H_1^3
                        +\frac{32}{3} H_1 H_{0,1}
                        -64 H_0 H_{0,-1}
                        \nonumber \\ &&
                        +32 H_0 H_{0,0,1}
                        +128 H_{0,0,-1}
                        -96 H_{0,0,0,1}
                        +\frac{64}{3} H_1 \zeta_2
                        +\frac{192}{5} \zeta_2^2
                        +64 H_0 \zeta_3
                        \nonumber \\ &&
                        -\frac{4}{9} \big(25+36 H_0\big) H_1^2
                \biggr)
                -\frac{2}{9} (1794-1777 x)
                +\biggl(
                         \frac{4}{27} (43-1949 x)
                         \nonumber \\ &&
                        +\frac{32 (1+x) (1-5 x)}{x} H_{-1}
                \biggr) H_0
                +\frac{2}{9} (400+x) H_0^2
                +\frac{4}{3} (26+5 x) H_0^3
                \nonumber \\ &&
                -\biggl(
                         \frac{8}{27} (470-493 x)
                        -\frac{16}{3} (73-63 x) H_0
                \biggr) H_1
                -\biggl(
                         \frac{16}{9} (247-176 x)
                         \nonumber \\ &&
                        -48 (4+3 x) H_0
                \biggr) H_{0,1}
                -\frac{32 (1+x) (1-5 x)}{x} H_{0,-1}
                -96 (4+3 x) H_{0,0,1}
                \nonumber \\ &&
                -\biggl(
                         \frac{16}{9} (44-13 x)
                        +\frac{32}{3} (10+x) H_0
                \biggr) \zeta_2
                +64 (1+7 x) \zeta_3
        \biggr)
        \nonumber \\ &&
        +\textcolor{blue}{C_A} 
        \biggl(
                 p_{gq} 
                \biggl(
                        \frac{16}{3} H_0^2 H_1
                        -8 H_0(x) H_1^2
                        +\frac{8}{9} H_1^3
                        -\frac{64}{3} H_0 H_{0,1}
                        -\frac{256}{3} H_{0,0,-1}
                        \nonumber \\ &&
                        -\frac{16}{3} H_1 \zeta_2
                \biggr)
                -\frac{8}{27} (2215-2207 x)
                -\biggl(
                         \frac{8}{27} (569+1550 x)
                         \nonumber \\ &&
                        +\frac{32 (3+2 x) (1-4 x)}{3 x} H_{-1}
                        -\frac{64}{3} (2+x) H_{-1}^2
                \biggr) H_0
                +\biggl(
                        \frac{4}{9} (32+137 x)
                        \nonumber \\ &&
                        -\frac{32}{3} (2+x) H_{-1}
                \biggr) H_0^2
                -\frac{8}{9} (2+x) H_0^3
                +\biggl(
                         \frac{8}{27} (32-229 x)
                        +\frac{32}{9} (1+4 x) H_0
                \biggr) H_1
                \nonumber \\ &&
                +\frac{4}{9} (134-121 x) H_1^2
                -\biggl(
                         \frac{16}{9} (16-17 x)
                        +\frac{64}{3} (2+x) H_{-1}
                \biggr) H_{0,1}
                \nonumber \\ &&
                +\biggl(
                         \frac{32 (3+2 x) (1-4 x)}{3 x}
                        +\frac{32}{3} (10-3 x) H_0
                        -\frac{128}{3} (2+x) H_{-1}
                \biggr) H_{0,-1}
                \nonumber \\ &&
                +\frac{32}{3} (2-5 x) H_{0,0,1}
                +\frac{128}{3} (1+x) H_{0,1,1}
                +\frac{64}{3} (2+x) H_{0,1,-1}
                +\frac{64}{3} (2+x) H_{0,-1,1}
                \nonumber \\ &&
                +\frac{128}{3} (2+x) H_{0,-1,-1}
                +\biggl(
                         \frac{16}{9} (74-25 x)
                        +\frac{64}{3} (5+x) H_0
                        +\frac{128}{3} (2+x) H_{-1}
                \biggr) \zeta_2
                \nonumber \\ &&
                +\frac{32}{3} (19-15 x) \zeta_3
        \biggr)
        +\textcolor{blue}{N_F} \textcolor{blue}{T_F} 
        \biggl(
                 \frac{64}{27} (40-23 x)
                +\frac{32}{9} (4+x) H_1
        \biggr)
        \nonumber \\ &&
        +p_{gq} \textcolor{blue}{T_F} 
        \biggl(
                \frac{992}{27}
                -\frac{16}{3} \textcolor{blue}{N_F} H_1^2
        \biggr)
\Biggr]
+\textcolor{blue}{T_F} 
\biggl[
        -\frac{16}{243} (862-485 x)
        \nonumber \\ &&
        +\textcolor{blue}{N_F}
        \biggl(
                 \frac{128}{243} (161-67 x)
                -\frac{64}{27} (6-5 x) H_1
                -\frac{32}{27} (4+x) H_1^2
        \biggr)
        -\frac{32}{27} (6-5 x) H_1
        \nonumber \\ &&
        -\frac{16}{27} (4+x) H_1^2
\biggr]
+p_{gq} \textcolor{blue}{T_F} 
\biggl[
        \textcolor{blue}{N_F} 
        \biggl(
                \frac{32}{27} H_1^3
                -\frac{256}{9} \zeta_3
        \biggr)
        +\frac{16}{27} H_1^3
        +\frac{448}{9} \zeta_3
\biggr]
\nonumber \\ &&
+\textcolor{blue}{C_A} 
\biggl[
        16 B_4 x
        -\frac{8}{243} (30001-31508 x)
        +p_{gq} 
        \biggl(
                -\frac{32}{3} B_4
                +\frac{8}{27} H_0^3 H_1
                -\frac{32}{9} H_0^2 H_1^2
                \nonumber \\ &&
                +\frac{16}{9} H_0 H_1^3
                -\frac{10}{27} H_1^4
                +\biggl(
                        \frac{64}{9} H_0 H_1
                        +\frac{32}{9} H_1^2
                \biggr) H_{0,1}
                +\biggl(
                        -\frac{104}{9} H_0
                        \nonumber \\ &&
                        +\frac{64}{9}  H_1
                \biggr) H_0 H_{0,-1}
                +\frac{32}{9} H_1 H_{0,0,1}
                -\frac{128}{9} H_1 H_{0,0,-1}
                -16 H_1 H_{0,1,1}
                +\frac{8}{9} H_1^2 \zeta_2
                \nonumber \\ &&
                -\frac{224}{9} H_1 \zeta_3
        \biggr)
        +\biggl(
                -\frac{8}{243} (4960+25273 x)
                +\frac{16 \big(9+173 x+70 x^2+12 x^3\big)}{27 x} H_{-1}
                \nonumber \\ &&
                +\frac{8 \big(-27+58 x+5 x^2\big)}{27 x} H_{-1}^2
                +\frac{176}{27} (2+x) H_{-1}^3
        \biggr) H_0
        \nonumber \\ &&
        +\biggl(
                 \frac{4}{81} \big(973+2677 x-72 x^2\big)
                +\frac{4 \big(45-238 x-203 x^2\big)}{27 x} H_{-1}
                \nonumber \\ &&
                -\frac{4}{3} (2+x) H_{-1}^2
        \biggr) H_0^2
        +\biggl(
                 \frac{152}{81} (3-2 x)
                +\frac{8}{27} (2+x) H_{-1}
        \biggr) H_0^3
        +\frac{4}{27} (4+x) H_0^4
        \nonumber \\ &&
        -\biggl(
                 \frac{8}{243} (980-1531 x)
                +\biggl(
                         \frac{4 \big(243-2488 x+1967 x^2\big)}{81 x}
                         \nonumber \\ &&
                        +\frac{16 (1-x) (1+5 x)}{x} H_{-1}
                \biggr) H_0
                +\frac{4 \big(-36-38 x+61 x^2\big)}{27 x} H_0^2
        \biggr) H_1
        \nonumber \\ &&
        -\biggl(
                 \frac{8}{81} (331-344 x)
                +\frac{4 \big(27+218 x-205 x^2\big)}{27 x} H_0
        \biggr) H_1^2
        -\frac{8}{81} (136-143 x) H_1^3
        \nonumber \\ &&
        +\biggl(
                \frac{4 \big(243-2408 x+1735 x^2\big)}{81 x}
                -\biggl(
                         \frac{16 \big(18-22 x-373 x^2\big)}{27 x}
                         \nonumber \\ &&
                        +\frac{64}{9} (2+x) H_{-1}
                \biggr) H_0
                +\frac{8}{9} (14+9 x) H_0^2
                +\frac{8 (1-x) (3+53 x)}{3 x} H_1
                \nonumber \\ &&
                -\frac{16 \big(27-52 x-119 x^2\big)}{27 x} H_{-1}
                -\frac{32}{9} (2+x) H_{-1}^2
                +\frac{32}{9} (22+7 x) H_{0,-1}
        \biggr) H_{0,1}
        \nonumber \\ &&
        +\frac{32}{9} (9+13 x) H_{0,1}^2
        +\biggl(
                -\frac{16 \big(9+173 x+70 x^2+12 x^3\big)}{27 x}
                \nonumber \\ &&
                -\biggl(
                         \frac{8 \big(117-238 x-495 x^2\big)}{27 x}
                        -\frac{16}{3} (2+x) H_{-1}
                \biggr) H_0
                +\frac{16 (1-x) (1+5 x)}{x} H_1
                \nonumber \\ &&
                +\frac{16 \big(27-58 x-5 x^2\big)}{27 x} H_{-1}
                -\frac{176}{9} (2+x) H_{-1}^2
        \biggr) H_{0,-1}
        +\frac{32}{3} x H_{0,-1}^2
        \nonumber \\ &&
        +\frac{32 \big(1-5 x^2\big)}{x} H_0 H_{-1,1}
        +\biggl(
                 \frac{8 \big(36-94 x-1775 x^2\big)}{27 x}
                -\frac{16}{9} (22+49 x) H_0
                \nonumber \\ &&
                +\frac{160}{9} (2+x) H_{-1}
        \biggr) H_{0,0,1}
        +\biggl(
                 \frac{8 \big(189-238 x-787 x^2\big)}{27 x}
                +\frac{16}{9} (50-47 x) H_0
                \nonumber \\ &&
                -\frac{16}{3} (2+x) H_{-1}
        \biggr) H_{0,0,-1}
        -\biggl(
                 \frac{8 \big(27+622 x-698 x^2\big)}{27 x}
                +\frac{32}{9} (8+15 x) H_0
                \nonumber \\ &&
                -\frac{128}{9} (2+x) H_{-1}
        \biggr) H_{0,1,1}
        +\biggl(
                 \frac{16 \big(27-52 x-119 x^2\big)}{27 x}
                -\frac{32}{9} (18+5 x) H_0
                \nonumber \\ &&
                +\frac{64}{9} (2+x) H_{-1}
        \biggr) H_{0,1,-1}
        -\biggl(
                 \frac{16 \big(27+52 x-151 x^2\big)}{27 x}
                +\frac{32}{9} (18-13 x) H_0
                \nonumber \\ &&
                -\frac{64}{9} (2+x) H_{-1}
        \biggr) H_{0,-1,1}
        -\biggl(
                 \frac{16 \big(27-58 x-5 x^2\big)}{27 x}
                +\frac{16}{3} (2+5 x) H_0
                \nonumber \\ &&
                -\frac{352}{9} (2+x) H_{-1}
        \biggr) H_{0,-1,-1}
        +\frac{16}{9} (30+137 x) H_{0,0,0,1}
        -\frac{16}{9} (74-101 x) H_{0,0,0,-1}
        \nonumber \\ &&
        -\frac{128}{3} (2+x) H_{0,0,1,1}
        -\frac{160}{9} (2+x) H_{0,0,1,-1}
        -\frac{32}{9} (10+41 x) H_{0,0,-1,1}
        \nonumber \\ &&
        +\frac{16}{3} (2+x) H_{0,0,-1,-1}
        +\frac{32}{9} (4-17 x) H_{0,1,1,1}
        -\frac{128}{9} (2+x) H_{0,1,1,-1}
        \nonumber \\ &&
        -\frac{128}{9} (2+x) H_{0,1,-1,1}
        -\frac{64}{9} (2+x) H_{0,1,-1,-1}
        -\frac{32}{9} (10+19 x) H_{0,-1,0,1}
        \nonumber \\ &&
        -\frac{128}{9} (2+x) H_{0,-1,1,1}
        -\frac{64}{9} (2+x) H_{0,-1,1,-1}
        -\frac{64}{9} (2+x) H_{0,-1,-1,1}
        \nonumber \\ &&
        -\frac{352}{9} (2+x) H_{0,-1,-1,-1}
        +\biggl(
                \frac{16}{81} \big(499+58 x+36 x^2\big)
                -\biggl(
                         \frac{64}{27} (4-43 x)
                         \nonumber \\ &&
                        +\frac{64}{9} (2+x) H_{-1}
                \biggr) H_0
                -\frac{64}{9} (1+2 x) H_0^2
                -\frac{16 \big(27+199 x-206 x^2\big)}{27 x} H_1
                \nonumber \\ &&
                -\frac{16 \big(27-139 x-86 x^2\big)}{27 x} H_{-1}
                +\frac{40}{3} (2+x) H_{-1}^2
                -\frac{16}{9} (50+31 x) H_{0,1}
                \nonumber \\ &&
                +\frac{16}{9} (26-11 x) H_{0,-1}
        \biggr) \zeta_2
        +\frac{16}{45}(365-413 x) \zeta_2^2
        +\biggl(
                \frac{20}{27} (326+403 x)
                \nonumber \\ &&
                +\frac{128}{9} (13+3 x) H_0
                -\frac{112}{3} (2+x) H_{-1}
        \biggr) \zeta_3
\biggr]
+\textcolor{blue}{C_F} 
\biggl[
        -32 B_4 x
        \nonumber \\ &&
        -\frac{1}{162} (316242-278981 x)
        +p_{gq} 
        \biggl(
                \frac{64}{3} B_4
                +\frac{8}{3} H_0^3 H_1
                +\frac{16}{9} H_0^2 H_1^2
                +\frac{32}{9} H_0 H_1^3
                \nonumber \\ &&
                +\frac{10}{27} H_1^4
                +\biggl(
                        \frac{64}{9} H_0 H_1
                        -\frac{16}{9} H_1^2
                \biggr) H_{0,1}
                +\frac{64}{3} H_0^2 H_{0,-1}
                -\frac{64}{9} H_1 H_{0,0,1}
                \nonumber \\ &&
                -\biggl(
                         \frac{128}{9} H_0
                        -\frac{32}{9} H_1
                \biggr) H_{0,1,1}
                +64 H_0 H_{0,-1,1}
                +64 H_0 H_{0,0,0,1}
                +\frac{128}{9} H_{0,0,1,1}
                \nonumber \\ &&
                -\frac{112}{9} H_{0,1,1,1}
                -256 H_{0,0,0,0,1}
                +\biggl(
                        -\frac{16}{9} H_0^2
                        -\frac{128}{9} H_0 H_1
                        -\frac{80}{9} H_1^2
                        -32 H_{0,-1}
                \biggr) \zeta_2
                \nonumber \\ &&
                +\frac{384}{5} H_0 \zeta_2^2
                +\biggl(
                        \frac{128}{3} H_0^2
                        +32 H_1
                \biggr) \zeta_3
                +256 \zeta_5
        \biggr)
        +\biggl(
                -\frac{4}{243} (54424+48499 x)
                \nonumber \\ &&
                -\frac{16 (1+x) \big(-5-13 x+4 x^2\big)}{9 x} H_{-1}
        \biggr) H_0
        -\biggl(
                 \frac{2}{81}\big(3373+55 x-144 x^2\big)
                 \nonumber \\ &&
                +\frac{32 (1+x) (1-5 x)}{3 x} H_{-1}
        \biggr) H_0^2
        +\frac{14}{81} (86-55 x) H_0^3
        +\frac{4}{27} (13-5 x) H_0^4
        \nonumber \\ &&
        +\biggl(
                 \frac{16}{243} (2227-1634 x)
                -\frac{16 \big(3+5 x-10 x^2\big)}{9 x} H_0^2
                +\biggl(
                        -\frac{16 \big(27-367 x+182 x^2\big)}{27 x}
                        \nonumber \\ &&
                        +\frac{32 (1-x) (1+5 x)}{x} H_{-1}
                \biggr) H_0
        \biggr) H_1
        +\biggl(
                 \frac{8}{81} (652-545 x)
                -\frac{8}{27} (62-x) H_0
        \biggr) H_1^2
        \nonumber \\ &&
        +\frac{40}{81} (5-4 x) H_1^3
        +\biggl(
                \frac{16 \big(81-818 x+220 x^2\big)}{81 x}
                +\frac{16 \big(18+428 x-655 x^2\big)}{27 x} H_0
                \nonumber \\ &&
                -\frac{8}{9} (38+5 x) H_0^2
                +\frac{16}{27} (46-17 x) H_1
                +\frac{32 (1+x) (1-5 x)}{x} H_{-1}
                \nonumber \\ &&
                -64 (2+x) H_{0,-1}
        \biggr) H_{0,1}
        -\biggl(
                 \frac{16 (1+x) \big(5+13 x-4 x^2\big)}{9 x}
                -\frac{64 \big(2-4 x-13 x^2\big)}{3 x} H_0
                \nonumber \\ &&
                +\frac{32 (1-x) (1+5 x)}{x} H_1
        \biggr) H_{0,-1}
        -\frac{64 \big(1-5 x^2\big)}{x} H_0 H_{-1,1}
        \nonumber \\ &&
        -\biggl(
                 \frac{16 \big(18+806 x-1249 x^2\big)}{27 x}
                -\frac{16}{9} (206+143 x) H_0
        \biggr) H_{0,0,1}
        \nonumber \\ &&
        -\biggl(
                 \frac{64 \big(3-4 x-21 x^2\big)}{3 x}
                +\frac{128}{3} (4-3 x) H_0
        \biggr) H_{0,0,-1}
        +\frac{32}{27} (11+26 x) H_{0,1,1}
        \nonumber \\ &&
        -\biggl(
                 \frac{32 (1+x) (1-5 x)}{x}
                -64 (2+x) H_0
        \biggr) H_{0,1,-1}
        +\frac{32 (1-x) (1+5 x)}{x} H_{0,-1,1}
        \nonumber \\ &&
        -\frac{592}{9} (14+11 x) H_{0,0,0,1}
        +256 (1-x) H_{0,0,0,-1}
        +256 x H_{0,0,-1,1}
        +128 x H_{0,-1,0,1}
        \nonumber \\ &&
        -\biggl(
                 \frac{16}{81} \big(121-326 x+36 x^2\big)
                +\frac{16}{27} (20-x) H_0
                -\frac{16 \big(27+124 x-119 x^2\big)}{27 x} H_1
                \nonumber \\ &&
                -\frac{16 (1+x) (1-5 x)}{x} H_{-1}
                -\frac{32}{9} (22+7 x) H_{0,1}
        \biggr) \zeta_2
        +\frac{32}{45} (323+509 x) \zeta_2^2
        \nonumber \\ &&
        +\biggl(
                 \frac{8}{27} (986-1861 x)
                +\frac{32}{9} (74+71 x) H_0
        \biggr) \zeta_3
\biggr]
\Biggr\}.
\end{eqnarray}
For the polarized massive operator matrix element the same set of 
harmonic polylogarithms contribute as in the unpolarized case.
The full set is reads: 
\begin{eqnarray}
&&
H_0,
H_{-1},
H_1,
H_{0,1}
H_{0,-1},
H_{-1, 1},
H_{0, 0, 1}
H_{0, 0, -1},
H_{0, 1, 1},
H_{0, -1, -1},
H_{0, 1, -1}.
H_{0, -1, 1,}, 
H_{0, 0, 0, 1}, 
H_{0, 0, 0, -1},
\nonumber\\ &&
H_{0, 0, 1, 1},
H_{0, 0, -1, -1},
H_{0, 0, -1, 1},
H_{0, 0, 1, -1},
H_{0, -1, 0, 1},
H_{0, 1, 1, 1}, 
H_{0, -1, -1, -1},
H_{0, -1, -1, 1},
H_{0, 1, -1, -1},
\nonumber\\ &&
H_{0, -1, 1, -1},
H_{0, -1, 1, 1},
H_{0, 1, -1, 1},
H_{0, 1, 1, -1},
H_{0, 0, 0, 0, 1}.
\end{eqnarray}
Methods and programs for the numerical evaluation of harmonic polylogarithms 
are given in~\cite{NUMERIC}.

At 2-loop order the OME is not altered when changing the renormalization scheme
of the heavy quark.
The additional terms when changing from the on-shell scheme to the $\MS$ scheme 
for $m = \bar{m}$ are given by:
\begin{eqnarray}
A_{gq,Q}^{(3),\MS}(N) &=&  A_{gq,Q}^{(3)}(N)  
- 32 a_s^3 \textcolor{blue}{C_F^2 T_F} \bar{p}_{gq} \Biggl[
\ln^2\left(\frac{{m}^2}{\mu^2}\right)  
+ \ln\left(\frac{{m}^2}{\mu^2}\right) \left(\frac{N-2}{3 (N+1)} - S_1\right) 
\nonumber\\ &&
- \frac{4 (2 + 5 N)}{9 (N+1)} + \frac{4}{3} S_1 
\Biggr] 
\end{eqnarray}
and
\begin{eqnarray}
A_{gq,Q}^{(3),\MS}(x) &=&  A_{gq,Q}^{(3)}(x)  
- 32 a_s^3 \textcolor{blue}{C_F^2 T_F} p_{gq} \Biggl[
\ln^2\left(\frac{{m}^2}{\mu^2}\right) - \ln\left(\frac{{m}^2}{\mu^2}\right) \left(\frac{4-5x}{3(2-x)} - 
\ln(1-x)\right)
\nonumber\\ &&
- \frac{4(4+x)}{9(2-x)} - \frac{4}{3} \ln(1-x)
\Biggr].
\end{eqnarray}
Here we identified the masses in the on-shell and $\MS$ scheme $m=\bar{m}$
to shorten the expressions.
It is straight forward to obtain the relation between the two
renormalization schemes \cite{MASS} while keeping also the
scale dependence.
The corresponding relation has been given e.g. in \cite{Klein:2009ig}.

The two-mass contributions to ${A}_{gq}^{(3)}(N)$ read
\begin{eqnarray}
      \lefteqn{{A}_{gq}^{(3),\sf two-mass}(N) = \textcolor{blue}{C_F T_F^2} (N+2)
      \biggl\{
            -\frac{128}{9 N (N+1)}
            \biggl[ 
                  L_2^3 + L_1^3 + \frac{3}{4} L_1 L_2 \bigl( L_2 + L_1 \bigr) 
            \biggr]}
            \nonumber \\ &&
            - \biggl[
                   \frac{64 (2+5 N)}{9 N (N+1)^2}
                  -\frac{64}{3 N (N+1)}  S_1    
            \biggr] 
            \bigl( L_2^2 + L_1^2 \bigr)
            + \biggl[
                  -\frac{64 \big(22+41 N+28 N^2\big)}{27 N (N+1)^3}
                  \nonumber \\ &&
                  +\frac{64 (2+5 N)}{9 N (N+1)^2}  S_1
                  -\frac{32}{3 N (N+1)} S_1^2
                  -\frac{32}{3 N (N+1)} S_2
                  -\frac{32}{N (N+1)} \zeta_2      
            \biggr]
            \bigl( L_2 + L_1 \bigr)
            \nonumber \\ &&
            -\frac{64 \big(98+369 N+408 N^2+164 N^3\big)}{81 N (N+1)^4}
            +\biggl(
                   \frac{64 \big(22+41 N+28 N^2\big)}{27 N (N+1)^3}
                   \nonumber \\ &&
                  +\frac{32}{3 N (N+1)} S_2
            \biggr) S_1
            -\frac{32 (2+5 N)}{9 N (N+1)^2}  S_1^2
            +\frac{32}{9 N (N+1)} S_1^3
            -\frac{32 (2+5 N)}{9 N (N+1)^2}  S_2
            \nonumber \\ &&
            +\frac{64}{9 N (N+1)}  S_3
            -\biggl(
                   \frac{64 (2+5 N)}{9 N (N+1)^2}
                  -\frac{64}{3 N (N+1)}  S_1
            \biggr) \zeta_2
            -\frac{128}{9 N (1+N)}  \zeta_3
      \biggr\}
      \nonumber \\ &&
      + {a}_{gq}^{(3),\sf two-mass}(N).
\nonumber\\ 
\end{eqnarray}
Correspondingly, the $x$-space result is given by
\begin{eqnarray}
      {A}_{gq}^{(3),\sf two-mass}(x) &=& \textcolor{blue}{C_F T_F^2}
      \biggl\{
             \frac{160}{9} (2-x) \bigl( L_1^3 + L_2^3 \bigr)
            - \frac{32}{3} (2-x) L_1 L_2 \bigl( L_1 + L_2 \bigr)
            \nonumber \\ &&
            + L_1^2 
            \biggl(
                   \frac{32 (4+x)}{9}
                  +48 (2-x) H_0(\eta )
                  -\frac{32}{3} (2-x) H_1
            \biggr)
            + L_2^2
            \biggl(
                  \frac{32 (4+x)}{9}
                  \nonumber \\ &&
                  -48 (2-x) H_0(\eta )
                  -\frac{32}{3} (2-x) H_1
            \biggr)
            + L_2 
            \biggl(
                   \frac{992}{27} (2-x)
                  -\frac{32}{3} (4+x) H_0(\eta )
                  \nonumber \\ &&
                  +32 (2-x) H_0^2(\eta )
                  +32 (2-x) H_0(\eta ) H_1
            \biggr)
            + L_1 
            \biggl(
                   \frac{992}{27} (2-x)
                   \nonumber \\ &&
                  +\frac{32}{3} (4+x) H_0(\eta )
                  +32 (2-x) H_0^2(\eta )
                  -32 (2-x) H_0(\eta ) H_1
            \biggr)
            \nonumber \\ &&
            +\frac{\big(1+\sqrt{\eta }\big)^2 (2-x)T_3}{6 \eta ^{3/2}}
            \biggl(
                   H_0(\eta )^2 
 H_{-1}\big(\sqrt{\eta }\big)
                  - 4 H_0(\eta ) H_{0,-1}\big(\sqrt{\eta }\big)
                  \nonumber \\ &&
                  + 8 H_{0,0,-1}\big(\sqrt{\eta }\big)
            \biggr) 
            +\frac{\big(1-\sqrt{\eta }\big)^2 (2-x) T_4}{6 \eta ^{3/2}} 
            \biggl(
                  H_0^2(\eta ) H_1\big(\sqrt{\eta }\big)
                  \nonumber \\ &&
                  - 4 H_0(\eta ) H_{0,1}\big(\sqrt{\eta }\big)
                  + 8 H_{0,0,1}\big(\sqrt{\eta }\big)
            \biggr)
            +\frac{8 T_6}{243 \eta }
            \nonumber \\ &&
            +\frac{20 \big(2-2 \eta ^2-x+\eta ^2 x\big) H_0(\eta )}{3 \eta }
            +\frac{T_5}{3 \eta } H_0^2(\eta )
            -\frac{16}{9} (2-x) H_0^3(\eta )
            \nonumber \\ &&
            -\biggl(
                  \frac{64}{27} (6-5 x)
                  +\frac{64}{3} (2-x) H_0^2(\eta )
            \biggr) H_1
            -\frac{32}{27} (4+x) H_1^2
            \nonumber \\ &&
            +\frac{32}{27} (2-x) H_1^3
            -\frac{256}{9} (2-x) \zeta_3
      \biggr\}
\end{eqnarray}
with
\begin{eqnarray}
      T_3 &=& -10 \eta ^{3/2}+5 \eta ^2+42 \eta -10 \sqrt{\eta }+5 ,
      \\
      T_4 &=& 10 \eta ^{3/2}+5 \eta ^2+42 \eta +10 \sqrt{\eta }+5 ,
      \\
      T_5 &=& 5 \eta ^2 x-10 \eta ^2+50 \eta  x+28 \eta +5 x-10 ,
      \\
      T_6 &=& 405 \eta ^2 x-810 \eta ^2+1130 \eta  x-1828 \eta +405 x-810 .
\end{eqnarray}

\vspace{5mm}\noindent
{\bf Acknowledgment.}~
This project has received funding from the European Union's Horizon 2020 research and innovation programme 
under the Marie Sk\l{}odowska--Curie grant agreement No. 764850, SAGEX and from the Austrian Science Fund (FWF) 
grant SFB F50 (F5009-N15). We would like to thank S.~Moch for a discussion and M.~Steinhauser for providing 
the code  {\tt MATAD 3.0}.

\end{document}